\documentclass[twoside,twocolumn,amsmath,amssymb,bm,gensymb,9pt]{article}
\usepackage[super,sort&compress,comma]{natbib} 
\usepackage[version=3]{mhchem}
\usepackage[left=1.5cm, right=1.5cm, top=1.785cm, bottom=2.0cm]{geometry}
\usepackage{balance}
\usepackage{mathptmx}
\usepackage{sectsty}
\usepackage{graphicx} 
\usepackage{lastpage}
\usepackage[format=plain,justification=justified,singlelinecheck=false,font={stretch=1.125,small,sf},labelfont=bf,labelsep=space]{caption}
\usepackage{float}
\usepackage{fancyhdr}
\usepackage{fnpos}
\usepackage[english]{babel}
\addto{\captionsenglish}{%
  
}

\usepackage{array}
\usepackage{droidsans}
\usepackage{charter}
\usepackage[T1]{fontenc}
\usepackage[usenames,dvipsnames]{xcolor}
\usepackage{setspace}
\usepackage[compact]{titlesec}
\usepackage{hyperref}
\usepackage{cleveref}
\usepackage[toc,page]{appendix}

\usepackage{tabularx}

\definecolor{cream}{RGB}{222,217,201}
\definecolor{BLUE}{RGB}{0,0,1}
\begin{document}

\pagestyle{fancy}
\thispagestyle{plain}
\fancypagestyle{plain}{
\renewcommand{\headrulewidth}{0pt}
}

\makeFNbottom
\makeatletter
\renewcommand\LARGE{\@setfontsize\LARGE{15pt}{17}}
\renewcommand\Large{\@setfontsize\Large{12pt}{14}}
\renewcommand\large{\@setfontsize\large{10pt}{12}}
\renewcommand\footnotesize{\@setfontsize\footnotesize{7pt}{10}}
\makeatother

\renewcommand{\thefootnote}{\fnsymbol{footnote}}
\renewcommand\footnoterule{\vspace*{1pt}%
\color{cream}\hrule width 3.5in height 0.4pt \color{black}\vspace*{5pt}} 
\setcounter{secnumdepth}{5}

\makeatletter 
\renewcommand\@biblabel[1]{#1}            
\renewcommand\@makefntext[1]%
{\noindent\makebox[0pt][r]{\@thefnmark\,}#1}
\makeatother 
\renewcommand{\figurename}{\small{Fig.}~}
\sectionfont{\sffamily\Large}
\subsectionfont{\normalsize}
\subsubsectionfont{\bf}
\setstretch{1.125} 
\setlength{\skip\footins}{0.8cm}
\setlength{\footnotesep}{0.25cm}
\setlength{\jot}{10pt}
\titlespacing*{\section}{0pt}{4pt}{4pt}
\titlespacing*{\subsection}{0pt}{15pt}{1pt}

\fancyfoot{}
\fancyhead{}
\renewcommand{\headrulewidth}{0pt} 
\renewcommand{\footrulewidth}{0pt}
\setlength{\arrayrulewidth}{1pt}
\setlength{\columnsep}{6.5mm}
\setlength\bibsep{1pt}

\makeatletter 
\newlength{\figrulesep} 
\setlength{\figrulesep}{0.5\textfloatsep} 

\newcommand{\topfigrule}{\vspace*{-1pt}%
\noindent{\color{cream}\rule[-\figrulesep]{\columnwidth}{1.5pt}} }

\newcommand{\botfigrule}{\vspace*{-2pt}%
\noindent{\color{cream}\rule[\figrulesep]{\columnwidth}{1.5pt}} }

\newcommand{\dblfigrule}{\vspace*{-1pt}%
\noindent{\color{cream}\rule[-\figrulesep]{\textwidth}{1.5pt}} }

\makeatother

\twocolumn[
  \begin{@twocolumnfalse}
{}
\par
\vspace{1em}
\sffamily
\begin{tabular}{m{4.5cm} p{13.5cm} }

\includegraphics{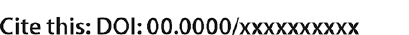} & \noindent\LARGE{\textbf{A Coarse-Grained Simulation Model for Colloidal Self-Assembly via Explicit Mobile Binders}} \\
\vspace{0.3cm} & \vspace{0.3cm} \\

 & \noindent\large{Gaurav Mitra,\textit{$^{a}$} Chuan Chang,\textit{$^{b}$} Angus McMullen,\textit{$^{c}$} Daniela Puchall,\textit{$^{a}$} Jasna Brujic,\textit{$^{c}$} and Glen M. Hocky \textit{$^{a,d,*}$} } \\

\includegraphics{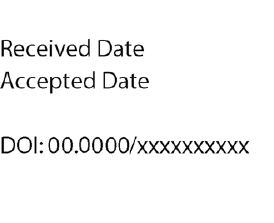} & \noindent\normalsize{
Colloidal particles with mobile binding molecules constitute a powerful platform
for probing the physics of self-assembly. Binding molecules are free to diffuse and rearrange on the surface, giving rise to spontaneous control over the number of droplet-droplet bonds, i.e., valence, as a function of the concentration of binders.
This type of valence control has been realized experimentally by tuning the interaction strength between DNA-coated emulsion droplets. 
Optimizing for valence two yields droplet polymer chains, termed `colloidomers', which have recently been used to probe the physics of folding.
To understand the underlying self-assembly mechanisms, here we present a coarse-grained 
molecular dynamics (CGMD) model to study the self-assembly of this class of systems using \textit{explicit representations of mobile binding sites}.
We explore how valence of assembled structures can be tuned through kinetic control in the strong binding limit. More specifically, 
we optimize experimental control parameters to obtain the highest yield of long linear colloidomer chains. 
Subsequently tuning the dynamics of binding and unbinding via a temperature-dependent model allows us to observe the heptamer chain collapse into all possible rigid structures, in good agreement with recent folding experiments. 
Our CGMD platform and dynamic bonding model (implemented as an open-source custom
plugin to HOOMD-Blue)  
reveal the molecular features governing the binding patch size and valence control, and opens the study of pathways in colloidomer folding. This model can therefore guide programmable design in experiments. 
} 

\end{tabular}

 \end{@twocolumnfalse} \vspace{0.6cm}
 ]


\renewcommand*\rmdefault{bch}\normalfont\upshape
\rmfamily
\section*{}
\vspace{-1cm}


\footnotetext{\textit{$^{a}$~Department of Chemistry, New York University, New York, New York, 10003, USA}}
\footnotetext{\textit{$^{b}$~Department of Physics, Cornell University, Ithaca, New York, 14853, USA.}}
\footnotetext{\textit{$^{c}$~Center for Soft Matter Research and Department of Physics, New York University, New
York, New York, 10003, USA.}}
\footnotetext{\textit{$^{d}$~Simons Center for Computational Physical Chemistry. New York University, New
York, New York, 10003, USA.}}
\footnotetext{\textit{$^*$~E-mail: hockyg@nyu.edu}}



\section{Introduction}
\label{sec:introduction}
Self-assembly of colloidal materials can create non-trivial and programmable structures with wide-ranging and tunable material properties. 
The spatio-temporal visualization of colloids renders them as useful model systems for probing the underlying physics behind assembly processes of molecular systems \cite{whitesides2002self,rogers2016using,cademartiri2015programmable,hueckel2021total}. 
The synthesis of colloidal particles with chemically or physically patterned solid surfaces---``patches''--- has been an active area of research due to the desire to control the bond valence and orientation, going beyond what can be achieved through isotropic interactions \cite{wang2012colloids,zhang2004self,hueckel2021total,kim2022advances}.
For example, a long-desired target is a diamond lattice, with an open structure that is difficult to achieve without an imposed tetrahedral symmetry \cite{he2020colloidal}.

The most common approach to engineering specific interactions between patches is to use complementary strands of DNA, whose interaction strength can be tuned by the length and specific sequence of nucleotides.
\cite{mirkin1996dna,alivisatos1996organization,biancaniello2005colloidal,nykypanchuk2008dna,jones2015programmable,rogers2016using}.
Each DNA duplex has an associated melting temperature ($T_{\textrm{melt}}$) and it is possible to employ multiple sets of complementary strands exhibiting different $T_{\textrm{melt}}$ to control the types of bonds present during an annealing protocol \cite{valignat2005reversible}. 

Self-assembly of patchy particles can be studied in simulations using short-range directional non-bonded interactions \cite{kern2003fluid} or using only pairwise interactions in combination with specific geometric constraints that prevent multiple bonding to the same patch \cite{zhang2004self,sciortino2007self,russo2009reversible,zhang2021sequence}.
Interactions due to many DNAs on a colloidal surface can also be computed using more detailed MD simulations \cite{obrien2016programming}, or through a mean-field approach \cite{varilly2012general}.
An alternative strategy to modeling explicit patches is to develop pair-potentials from inverse design principles, which have been successfully leveraged to produce systems that assemble with low valence, such as dimers or chains\cite{banerjee2019assembly,dijkstra2021predictive}.
 
An intriguing alternative to patchy particles of fixed valence is a system of particles coated with mobile adhesion molecules. 
In this case, it is possible for particles to ``choose'' their valence based on the number of available neighbors with complementary binding molecules, minimizing the total free energy of the system. 
Experimentally, using oil droplets in water provides a mobile interface on which the DNA linkers segregate into patches by diffusion to give rise to e.g., dimers when all the DNA is recruited into a single patch or droplet chains at higher concentrations of DNA where two patches per droplet are favored \cite{mcmullen2018freely}. Alternatively, solid colloids can be coated with fluid lipid bilayers 
\cite{van2013solid,feng2013specificity}. 
The mobility of the linkers broadens the melting temperature window of the DNA, facilitating equilibrium self-assembly \cite{xia2020linker,rogers2016using}. Particles with mobile binders can also serve as a physical mimic for biological adhesion, where cells use a variety of dynamic binding molecules to stick to each other and to surfaces \cite{bell1984cell}. To this end, biomimetic emulsion droplets have been successfully functionalized with adhesive mobile proteins, such as biotin-streptavidin complexes \cite{hadorn2012specific, pontani2012biomimetic}, cadherin ectodomains \cite{pontani2016cis}, or other ligand-binder pairs relevant for immunotherapy \cite{mbarek2015phagocytosis}. 
Early theoretical work expanding on the model of Ref.~\citenum{varilly2012general} predicts that such particles could have an equilibrium valence that depends on the number of available neighboring particles in the system\cite{angioletti2014mobile}.

Our previous work shows that monodisperse PDMS oil droplets functionalized with different flavors of single-stranded DNA on the surface can self-assemble into structures of tunable valence \cite{pontani2013immiscible,elbers2015bulk,feng2013specificity,zhang2017sequential,zhang2018multivalent,mcmullen2018freely}. Under conditions where DNA bonds are
reversible at room temperature, these systems achieve their equilibrium valence configuration. These results are predicted by a free-energy functional that takes into account the molecular properties of the system, including DNA binding strength, flexibility, steric repulsion, and concentration \cite{mcmullen2021dna}. Optimizing for valence two, the self-assembly of complementary DNA-coated droplets yields linear \textit{colloidomer} chains  \cite{mcmullen2018freely}. Further programming the secondary interactions along the chains offers a 
physical model system to probe the energy landscape of biopolymer folding, and for building small `foldamer' structures that can serve as the basis for larger scale assemblies \cite{coluzza2013design,coluzza2013sequence,mcmullen2022self}. 
 A complementary work demonstrated the formation of reconfigurable colloidal molecules using polydisperse droplets surrounded by ligands \cite{chakraborty2022self}.
 
 Motivated by these experimental studies, our work develops a coarse-grained molecular dynamics (CGMD) simulation model and framework to study the self-assembly of these colloidal chains with mobile binders, and their subsequent folding. 
 The crucial feature of our model is the use of \textit{explicit mobile linkers} with bonds between complementary binding partners.
 Prior work on these types of systems used implicit models of binding between neighboring droplets via the formation of adhesion patches, and some models included approximations to account for the dynamics of adhesion\cite{angioletti2014mobile,angioletti2016theory,bachmann2016bond,bachmann2016melting,mognetti2019programmable,sanchez2022kinetically}. 
 The use of explicit mobile binders allows us to test the underlying assumptions in a more realistic model, albeit at the cost of additional complexity.
 For example, our model explicitly shows how steric repulsion between binders (designed to mimic electrostatic repulsion between DNA strands) affects the adhesion patch size and the concentration of binders therein. These results in turn explain the overall valence distribution that results from the assembly process. 
 To demonstrate the capabilities of our model, we describe the parameters that optimize the formation of colloidal chains under kinetic control, and show that it exhibits the folding behavior of two-dimensional colloidal chains commensurate with what has been recently demonstrated experimentally \cite{mcmullen2022self}.
 
\section{Description of the model}
\label{sec:cgmodel}
\subsection{Coarse-Grained model for colloidal particles with explicit mobile binders}
\begin{figure}[!ht]
    \centering
\includegraphics[width=0.9\columnwidth]{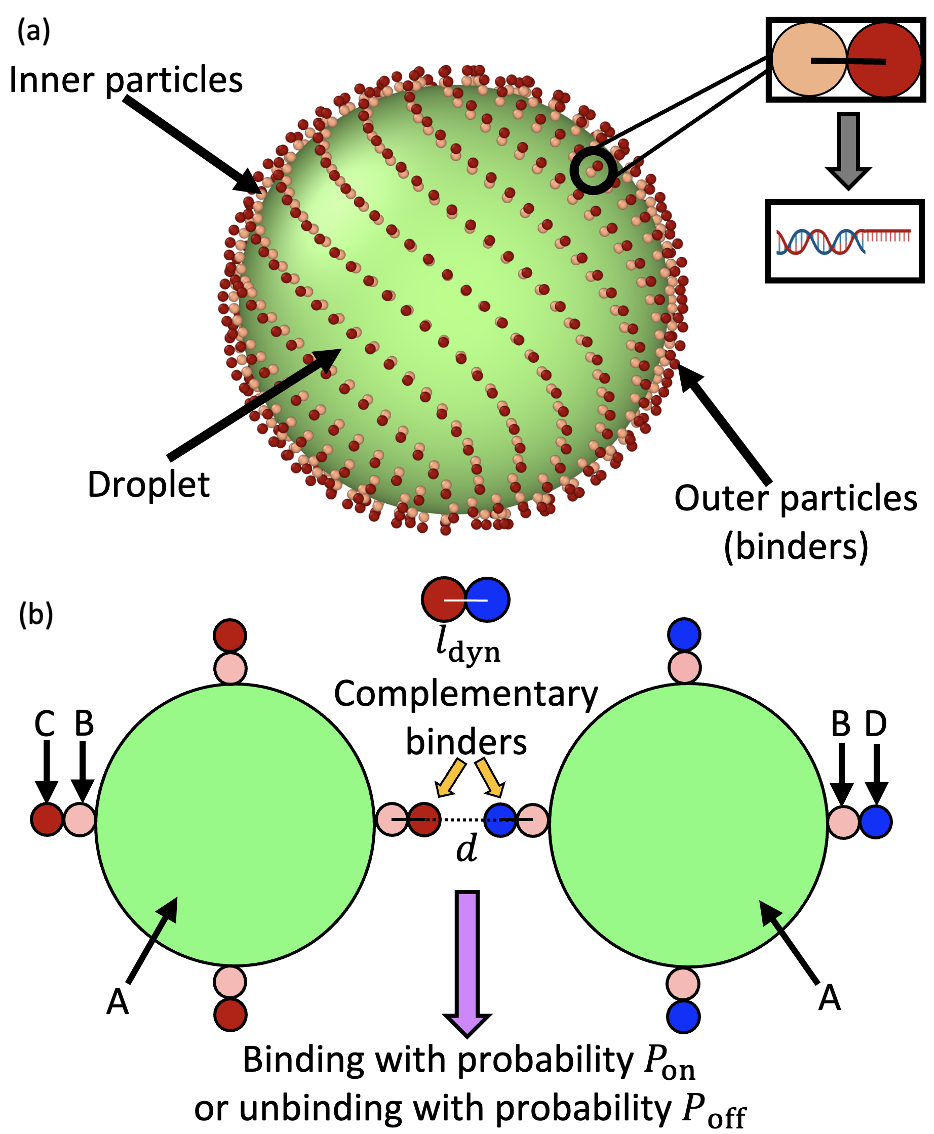}
    \caption{(a) The initial configuration of a droplet with binders adhered to its surface, arrayed in their initial ``Fibonacci'' structure. As shown in the inset, each binder consists of two constituent particles, which in the case of DNA corresponds to a spacer double-stranded sequence and a single stranded sequence which is available to bind to a complementary strand. 
    (b) A schematic showing dynamic binding between the outer binder particles of two droplets.
    Different particle types used in our python framework are schematically labeled A--D, with C and D representing a pair of complementary binders.}
    \label{fig:CGmodel}
\end{figure}

The central unit of our simulation model is a droplet, as shown in Fig.~\ref{fig:CGmodel}a.
Each droplet consists of a central spherical particle of radius $R$ (type A) with $N_b$ binders distributed on the surface. 
The positions of the binders are initialized in a ``Fibonacci'' arrangement to prevent overlap between particles in the initial configurations \cite{swinbank2006fibonacci}. 
Each binder is composed of two particles (types B and C) --- the outer particle is responsible for binding complementary partners, while the inner one is used to modulate excluded volume between binders. This pair of particles mimics the combination of the double-stranded tether and the single-stranded sticky-end DNA used in experiments \cite{mcmullen2018freely} (Fig.~\ref{fig:CGmodel}a). This configuration also allows us to apply an angular term to tune the propensity of the binder to stand vertically from the surface.
The binders diffuse on the surface \cite{feng2013specificity} due to a harmonic bond between the center of the droplet and that of the inner binder particle with a spring constant $k_{AB}$ and rest length $l_{AB}^0$.

The inner and outer particles in the binder are similarly held together by a harmonic bond with spring constant $k_{BC}$ and $l_{BC}=r_B+r_C$, the sum of their radii.
To have the binder stick outward from the droplet, the two binder particles are forced to align along the radial vector from the center of the droplet, by introducing a harmonic angular potential between the triplet of particles with parameters $k_{ABC}$ and rest angle  $\theta^0_{ABC}=180^\circ$.
In each case, the spring constants are chosen such that the thermal standard deviation $\sigma=\sqrt{k_\mathrm{B} T/k}$ (where $k_{\mathrm{B}}$ is the Boltzmann's constant) is a small fraction of the rest length or angle (see Tab.~\ref{tab:tablegeneral}).
As a trade-off between enforcing rigid bonds and using an infinitesimal MD time step, we chose to use spring constants where the standard deviation is $1-2\%$.

\subsection{Non-bonded interactions}
To prevent overlap between particles, we use a soft repulsion given by Eq.~\ref{equation:softV_nosmoothing}, which is applied between all particle types except between pairs of outer binder particles of complementary types (otherwise repulsion can prevent binding). 
By tuning the effective diameter of B particles, we tune the steric repulsion between adjacent binders, which corresponds to the case of altering the screening of the electrostatic interactions between DNA strands on the surface \cite{mcmullen2021dna}.

Our CGMD model can be used to study droplet interactions in three dimensions, but we implement 2D confinement for comparison with recent experiments. 
To replicate this quasi two-dimensional arrangement in our system, we confine droplets above and below using a force-shifted Lennard-Jones wall potential \cite{toxvaerd2011communication} on each droplet, between a fixed $z$-position and the center of the droplet A particle. 
The origins of the walls are given by $(0,0,2.5R)$ and $(0,0,-2.5R)$.

\subsection{Dynamic bonding model}
We model interactions between binders through covalent bonds. 
To do so, we develop a plugin to HOOMD-Blue \cite{anderson2008general} \cite{anderson2020hoomd} that builds upon a model for epoxy binding developed in Ref.~\citenum{thomas2018routine}.
Adhesive bonds form only between complementary outer binder particles of respective droplets, as shown in Fig.~\ref{fig:CGmodel}b.
In the simplest case, we have a mixture of droplets containing outer binder particles that are 100\% of types C and D, respectively. The model allows for individual droplets to contain mixtures of binder types, and there may be many more than two types, as designated by the user.
In this study, harmonic bonds are added with spring constant $k_{\mathrm{dyn}}$ and length $l_{\mathrm{dyn}}=r_C+r_D$, the sum of the radii of particles forming a bond. Here, we choose harmonic bonds, but any form of bond implemented in HOOMD-Blue could be used, since our algorithm only changes the bond table within the MD simulation, and does not compute or apply forces.\footnote{The Metropolis criterion employed for binding/unbinding described later requires knowing the energy of adding or removing a bond. At this time only a harmonic interaction is supported, although this can be trivially extended.}  

The effective free energy difference between a bound and an unbound state in a two-state binding reaction is given by 
\begin{equation}
\Delta G =k_{\mathrm{B}} T \ln\Bigg(\frac{k_\textrm{on}}{k_\textrm{off}}\Bigg) \equiv \varepsilon
\end{equation} 
Here, ${k_\textrm{on}}$ and ${k_\textrm{off}}$ are the rate constants for binding and unbinding, respectively. 
Below, we tune $\varepsilon$ to modulate the affinity between individual binders.

In Sec.~\ref{appendix:algorithm} we describe details of the algorithm for binding and unbinding, satisfying detailed balance. In addition, the temperature dependence is also described (Sec.~\ref{appendix:temperature}), in which the binding and unbinding rates are tuned as a function of temperature in such a way that the fraction of bound pairs at the equilibrium distance tends smoothly to zero at high $T$, with 50\% bound pairs at a specified melting temperature $T_\mathrm{melt}$.
\begin{figure*}[ht!]
    \centering
    \includegraphics[width=\textwidth]{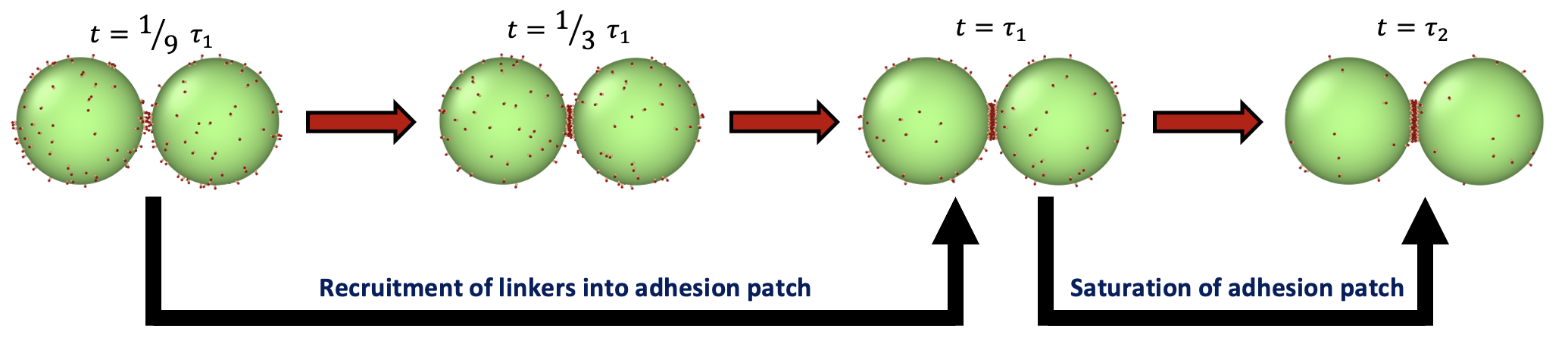}
    \caption{ Illustration of adhesion patch formation for a dimer of droplets. Patch formation under conditions with high binding affinity occurs via a process with two time scales ($\tau_1$ and $\tau_2$), one for recruitment of most linkers into a patch, and a second proceeding to saturation. The conditions for this particular simulation are $R=50$, $N_b=100$ and $\varepsilon=20.7$. See Sec.~\ref{appendix:convergence} and Tab.~\ref{tab:tabletimescales} for more details.}
    \label{fig:timescales}
\end{figure*}

\subsection{Comparison of model and experiment geometry}
This CGMD model is generic, while Fig.~\ref{fig:modeling} shows how two particles describe the experimental setup in Ref.~\citenum{mcmullen2018freely}.
The radius of one binder sphere corresponds to the sticky end of length $\sim5.1$nm, which is our reduced unit of length. 
Therefore, a droplet radius of $R=300$ corresponds to the droplet size of $\sim1530$nm used in Ref.~\citenum{mcmullen2018freely}.
Smaller particle sizes studied in Sec.~\ref{sec:results} are also used in experiments. 

Practically, our simulations are computationally limited to hundreds of binders per droplet. Therefore, we equate the scale of the simulations to experiments 
by matching the excluded volume of all binders to that of DNA surface coverage in the experiment. More specifically, the surface coverage, $p$ is defined as

\begin{equation}
p =  \frac{ 4\pi{r_{B}}^{2}N_{b} }{4\pi R^{2}}=N_b \left(\frac{r_B}{R} \right)^2.
\end{equation}

In the experiment, for droplets of radius $R=1530$ nm, an effective repulsive radius of DNA of 1.5nm \cite{ferrari1992scattering,koltover2000dna,mcmullen2021dna}, and an estimated $1\times10^3-2\times10^4$ DNA strands\cite{mcmullen2018freely,mcmullen2021dna},
$p$ ranges from $\sim0.001-0.02$ or $0.1-2\%$ coverage. For many simulations below, we use $r_B=1$, $R=50$, and $N_b=100$, in which case $p=0.04$.
From this perspective, each binder plays the collective role of hundreds of DNA.

\section{Simulation methods}
\label{sec:methods}

MD simulations of droplets coated with mobile binders were performed using HOOMD-blue version 2.9.6 \cite{anderson2008general,anderson2020hoomd,glaser2015strong}. 
The Langevin integrator \cite{phillips2011pseudo} was used to integrate all particles forward in time. Two different values of the drag coefficient $\gamma$ were used: one for the droplets ($\gamma_\textrm{A}$) and the other for the binders ($\gamma_\textrm{binder}$).
\par
The equation of motion for each particle $i$ in Langevin dynamics \cite{leimkuhler2015molecular} is given by:
\begin{equation}
m_i \ddot{\vec{r_i}}\big(t\big) = \vec{F}_i -\gamma_i\dot{\vec{r_i}}\big(t\big) + \sqrt{2\gamma_i k_{\mathrm{B}}T}\vec{\eta}\big(t\big)
\label{equation:langevin}
\end{equation}
where,
$m_i$ is the mass of the particle, $k_{\mathrm{B}}$ is the Boltzmann constant, $\gamma_i$ is the drag coefficient,  $\dot{\vec{\boldsymbol{r_i}}}(t)$ is the velocity of the particle, $\vec{F_i}=-\nabla U_i$ is the force on particle $i$ derived from the total potential energy function of the system, and $\eta(t)$ is the delta-correlated random white noise, with zero mean and unit variance. 
We use the tree neighbor list \cite{howard2016efficient,howard2019quantized} to accelerate non-bonded calculations, and in the construction of our list of possible pairs to bond as described above.

\section{Results and Discussion}
\label{sec:results}

\subsection{Main objectives}
In this section, we optimize simulation conditions to robustly self-assemble long colloidomers, without branched structures. 
We explore both the molecular properties of the system (droplet radius, binder concentration, and binder interaction strength), and the experimental conditions for assembly (particle concentration and solution viscosity). The detailed parameters used for our MD simulations are listed in \cref{tab:tablegeneral,tab:tabledimertrimer,tab:tablelattice,tab:tablefolding}.
Our results indicate that kinetic factors can be rationally employed to target the desired outcome with high yield and fidelity for fixed-time experiments. We subsequently show that our model allows us to study the folding process for  self-assembled colloidomer chains.

\begin{figure*}[ht]
    \centering
    \includegraphics[width=\textwidth]{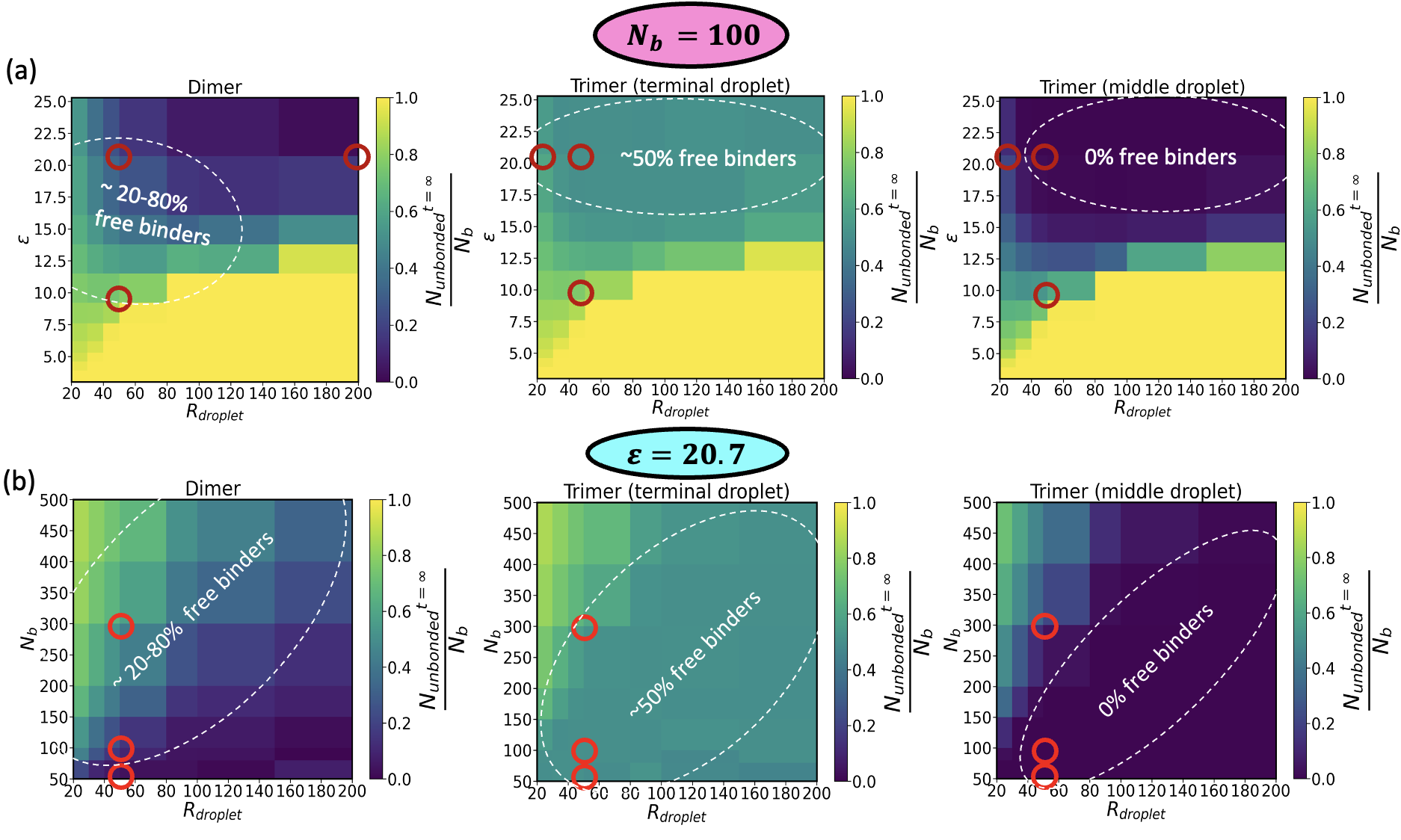}
    \caption{Heat maps showing the percentage of binders remaining on the surface of a droplet in a dimer geometry, and on the terminal or central droplets in a trimer geometry. Conditions predicted to be ``good''  for colloidomer assembly are indicated by dashed ovals, and selected conditions marked by red open circles are illustrated in Fig.~\ref{fig:dimertrimer_pictures}. (a) Fraction of remaining binders at fixed number of binders, varying droplet radius and bond strength.
    (b) Fraction of remaining binders at fixed high bond strength, varying number of binders and droplet radius.}
    \label{fig:dimertrimer_phasediagrams}
\end{figure*}

\subsection{Adhesion patch formation is a two-stage process for high bond strengths}
\label{sec:dimertrimer}
The formation of chains requires that each droplet has two contacting neighbors. 
Monomers first form dimers, after which they either combine with monomers to make trimers, or with other dimers to make tetramers. 
We therefore first probe the physical processes involved in forming a patch in a dimer or trimer configuration, and then   consider de novo assembly in Sec.~\ref{sec:lattice} .

Simulations of patch formation begin with dimers and trimers in an initial configuration with a single bond already formed between the droplets. Subsequently, the patches progressively grow until they reach steady-state. 
We find that patch formation (at intermediate and high binding affinities, $\varepsilon \ge 10$) happens in two stages, as illustrated in Fig.~\ref{fig:timescales} (see Fig.~\ref{fig:dimertrimer_convergence}). Fitting the fraction of unbound binders versus time with a double exponential function reveals two time scales, as shown in Sec.~\ref{appendix:convergence}. Initially, the fast time scale of recruitment of binders describes the formation of a stable adhesion patch ($\tau_1$), while the saturation of the patch is captured by a 1--2 orders of magnitude slower timescale ($\tau_2$). Tab.~\ref{tab:tabletimescales} reports values of $\tau_1$ and $\tau_2$ for some of these conditions.
 
\begin{figure*}[!ht]
    \centering
    \includegraphics[width=\textwidth]{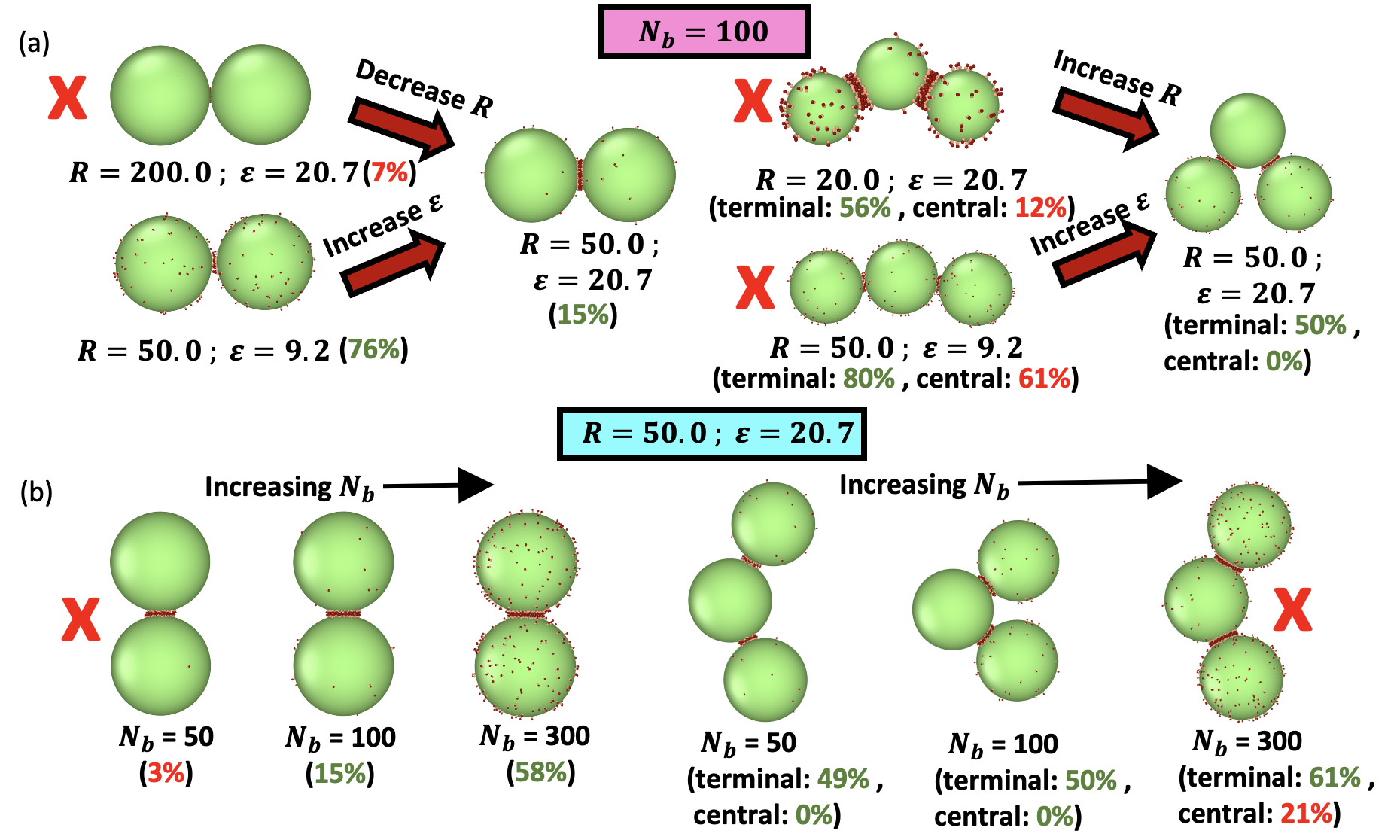}
    \caption{Illustrations showing the dimer and trimer geometries simulated in Fig.~\ref{fig:dimertrimer_phasediagrams}, and the ``molecular'' features of our droplet model that can be tuned to optimize for linear chains. Both results for dimers and trimers must be considered to predict the resulting polymerization reaction. (a) Varying $R$ and $\varepsilon$ at fixed number of binders $N_{b}=100$, with dimer on left and trimer on right. 
    (b) Varying the number of binders $N_b$ at fixed $R=50$ and $\varepsilon=20.7$. 
    In (a) and (b), the \% of free binders available (averaged over 10 independent runs) is indicated in parentheses beside each of these conditions for a droplet in a dimer as well as the terminal and central droplet(s) of a trimer. Based on the values of the \% of free binders available, the conditions which are not suitable for colloidomer assembly (almost all used up in dimer, too many remaining on central droplet in a trimer) are indicated by a red X.}
    \label{fig:dimertrimer_pictures}
\end{figure*}
\subsection{Optimizing bond strength, droplet size, and binder concentration for colloidomerization}
\label{sec:molecular_optimization}
Optimal conditions for colloidomerization require considering both the dimer and the trimer assembly, since we must find a condition where not all binders are exhausted in the dimer, where $\sim50\%$ of binders are available on a terminal droplet of a trimer, and where very few are left on the trimer middle droplet, preventing branching. Fig.~\ref{fig:dimertrimer_phasediagrams}a shows a systematic study of the self-assembly of droplets of varying $R$ and $\varepsilon$, for a fixed total number of binders $N_b=100$.  
The heat maps show the fraction of unbound binders in three configurations: a droplet-droplet dimer, a terminal droplet in a trimer, and a central droplet in a trimer.
Conditions for colloidomerization are satisfied in the region where the dashed ovals overlap. 
Later, we choose $R=50$, $\varepsilon=20.7$ and $N_b=100$ to demonstrate that this condition results in the robust assembly of colloidomers.

The number of binders in a patch at equilibrium is dictated by the free energy of patch formation, which can be considered as the difference in the chemical potential inside and outside the patch \cite{mcmullen2021dna}. 
The driving force for a binder to enter the adhesion patch is determined by the energy of forming individual bonds, $\varepsilon$, and opposed by steric repulsion between binders, the stretching of binders at the interface, as well as the loss of entropy as the binder motion is constrained in a patch.
As $\varepsilon$ increases, the fraction of free binders decreases monotonically in the case of both dimers and trimers until it reaches an asymptotic value that is limited by the steric repulsion between binders. 

Similarly, increasing droplet size in the strong binding limit recruits progressively more binders into the patch, since there is more space at the interface between larger droplets (Fig.~\ref{fig:dimertrimer_phasediagrams}a). 
For large $R$ in both dimers and trimers, we observe that the transition from no binding to all binders in the patch is very sharp as energy gain overtakes entropic losses without a penalty from steric repulsion. In contrast, for small radii, the recruitment of binders into the patch is more gradual with $\varepsilon$ due to crowding.  

In addition to binding enthalpy gain and crowding, there is also the entropic cost of patch formation, which increases with droplet size. Therefore, at intermediate values of $\varepsilon$ we observe a non-monotonic recruitment of binders into the patch as a function of droplet size. While crowding dominates in small droplets, entropic costs dominate in large droplets, giving rise to an optimal size for binder recruitment, e.g. $R=120$ for $\varepsilon=15$ in the dimer configuration. 

We can now consider the transition from dimer to trimer, where two adhesion patches are formed. 
Competition between the two adhesion patches results in an even split of binders on the middle droplet; therefore, even in cases where more than half of binders can pack into a patch, only half of the binders on each terminal droplet are exhausted. 
This situation occurs for higher $\varepsilon$, whereas for weaker binding, entropy dominates and many binders remain outside of the two adhesion patches on both the central and terminal droplets. 

In Fig.~\ref{fig:dimertrimer_phasediagrams}b we fix $\varepsilon=20.7$ at the goldilocks value from Fig.~\ref{fig:dimertrimer_phasediagrams}a, and explore the effect of varying binder surface coverage. 
In this strong binding regime, we eventually saturate the geometric limit set by the droplet size and binder repulsion (see also Fig.~\ref{fig:saturation}). 
If $N_b$ is increased above this limit, it simply results in additional free binders. 
For small droplets, we quickly reach the situation where not all binders on the middle droplet of a trimer can fit into two patches, but we see that for larger droplets there is a very wide tolerance for binder concentration which might be useful for colloidomerization. 

In summary, for certain parameters (including $\varepsilon=20.7, R=50, N_b=100$) we find that $<100\%$ of the binders are recruited for dimers, while in trimers patches contain exactly half of the binders due to the competition between neighbors. This is an optimal situation for the self-assembly of colloidomers, and we proceed to study assembly of these droplets in Sec.~\ref{sec:lattice}.

\begin{figure*}[!ht]
    \centering
    \includegraphics{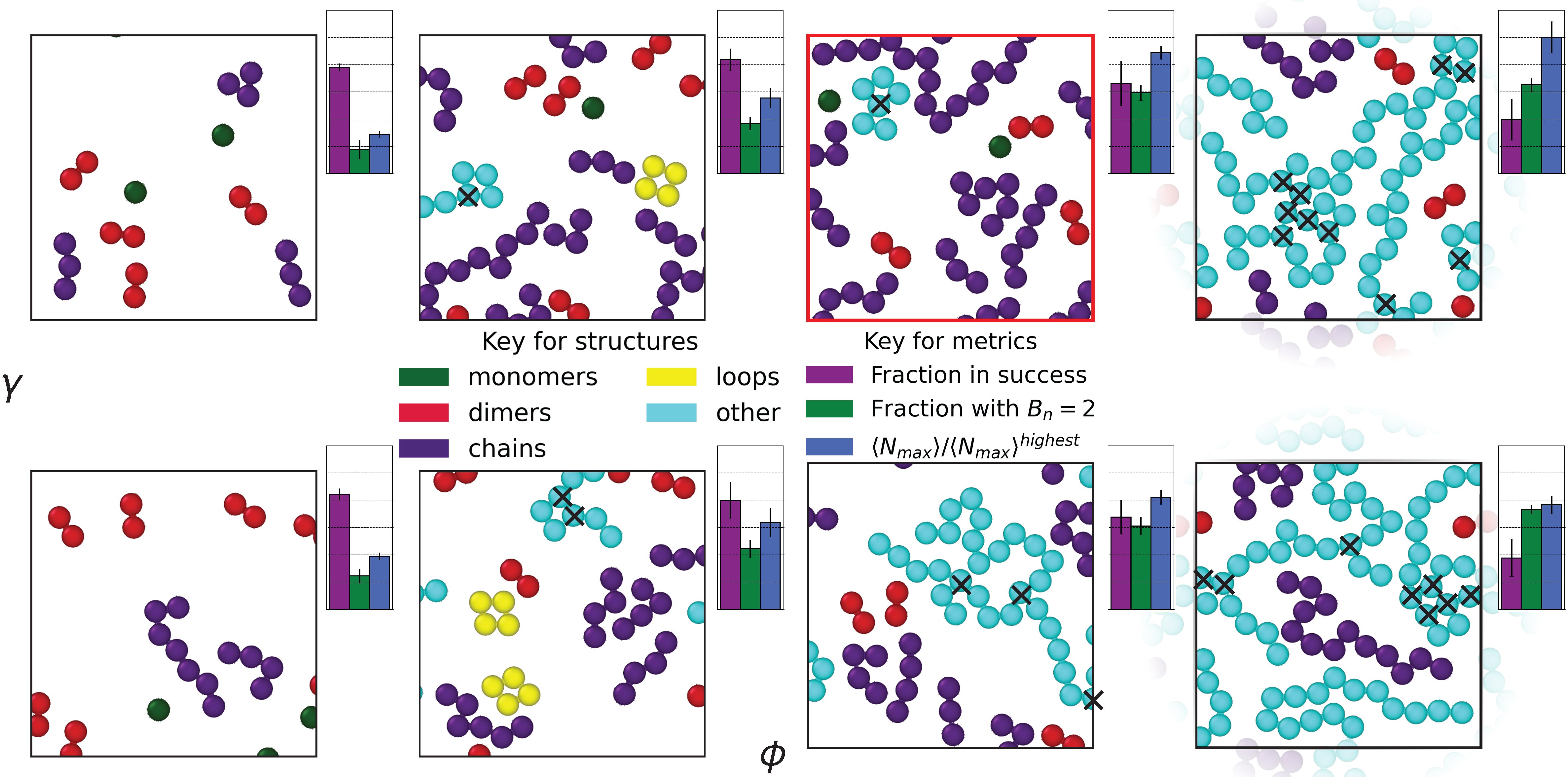}
    \caption{Effect of area fraction $\phi$ and the droplet drag coefficient $\gamma_A$ on self-assembly in a 1:1 mixture of 81 $R=50$ droplets containing $N_b=100$ complementary binders of type C and D (with one surplus droplet of C/D). $\varepsilon = 20.7$. Here, $\phi = 0.1,0.2,0.3,0.4$ (increasing from left to right) and $\gamma_A = 0.01,1.0$ (bottom and top). Every system is the same size, but each snapshot has identically sized field of view by area, meaning droplets are cropped at $\phi<0.4$.  Droplets are colored according to structure as shown in the key.
    Each condition is accompanied with bar charts representing quantities computed over 10 independent runs that can quantitatively help in collectively deciding the `winning condition' for colloidomer formation. These quantities are labeled in the second key, and described in detail in the main text, and for each quantity, optimal would be a larger bar.
    The few droplets that have valence 3 or higher are also marked with a `cross' in the representative configurations to reinforce that even the branched structures which are considered part of `errors' have long segments of droplets with valence=2. The condition we consider optimal, $\phi=0.3, \gamma_A=1.0$ is highlighted with a red box. 
    Snapshots at $\phi=0.4$ show periodic images to emphasize that structures are extended across the periodic boundaries.
    The distributions of structures obtained and of the colloidomer chain lengths for each of these 8 conditions are illustrated in Fig.~\ref{fig:histograms_finalstructures} and Fig.~\ref{fig:chainlength_histograms} respectively.}
    \label{fig:density_gamma_higheps}
\end{figure*}
\subsection{Illustrating a molecular recipe for colloidomers}

Fig.~\ref{fig:dimertrimer_pictures} illustrates scenarios that are predicted to be good or bad  for colloidomer assembly, as described above; full trajectories for these conditions are also shown in Movies M1 and M2.
Considering the dimer, we see that large droplets and high $\varepsilon$ allow almost all the binders to fit into a patch, which is detrimental for colloidomer assembly because it terminates the polymerization reaction. Decreasing droplet size to $R=50$ limits binders due to their steric repulsion in the patch, leaving just enough binders to seed a trimer with no remaining binders on the middle droplet, thus imposing the self-assembly of linear chains. Lowering the droplet size further or decreasing the binding strength 
leaves too many binders free on the trimer middle droplet at equilibrium, which would eventually lead to the self-assembly of branched colloidomers. 
Fig.~\ref{fig:dimertrimer_pictures}b illustrates that for a fixed droplet size and bond strength, the number of binders must be chosen such that dimers have free binders to grow the chain, while trimers have no free binders for branching via the middle droplet, as is the case for $N_b=100$. 
On the other hand, $N_b=50$ has almost all the binders ($97\%$) recruited in the case of the dimer, terminating colloidomer assembly. For $N_b=300$, the central droplet of a trimer has $21\%$ binders remaining on the surface, which would result in the branching of colloidomers.

\subsection{Kinetic optimization of colloidomerization}
\label{sec:lattice}
Beyond dimers and trimers, the self-assembly of chains requires further optimization of competing experimental timescales. 
Combining a 1:1 mixture of droplets containing complementary binders of particles type `C' and `D' mimics the experiments in Ref.~\citenum{mcmullen2018freely}. Tuning the density $\phi$ (area fraction) of droplets for a finite assembly time effectively controls the collision time between droplets and allows us to optimize the formation of long colloidomers out of equilibrium.    

Given that the droplets are undergoing simple diffusion, the collision time $\tau_\textrm{collision}\sim \langle l^2 \rangle $/D, where $l$ is the distance between droplets and $D$ is the diffusion constant. In two dimensions, $l^2 \sim \phi^{-1}$, such that the collision time is inversely proportional to the droplet density.

\begin{figure}[t]
    \centering
    \includegraphics[width=\columnwidth]{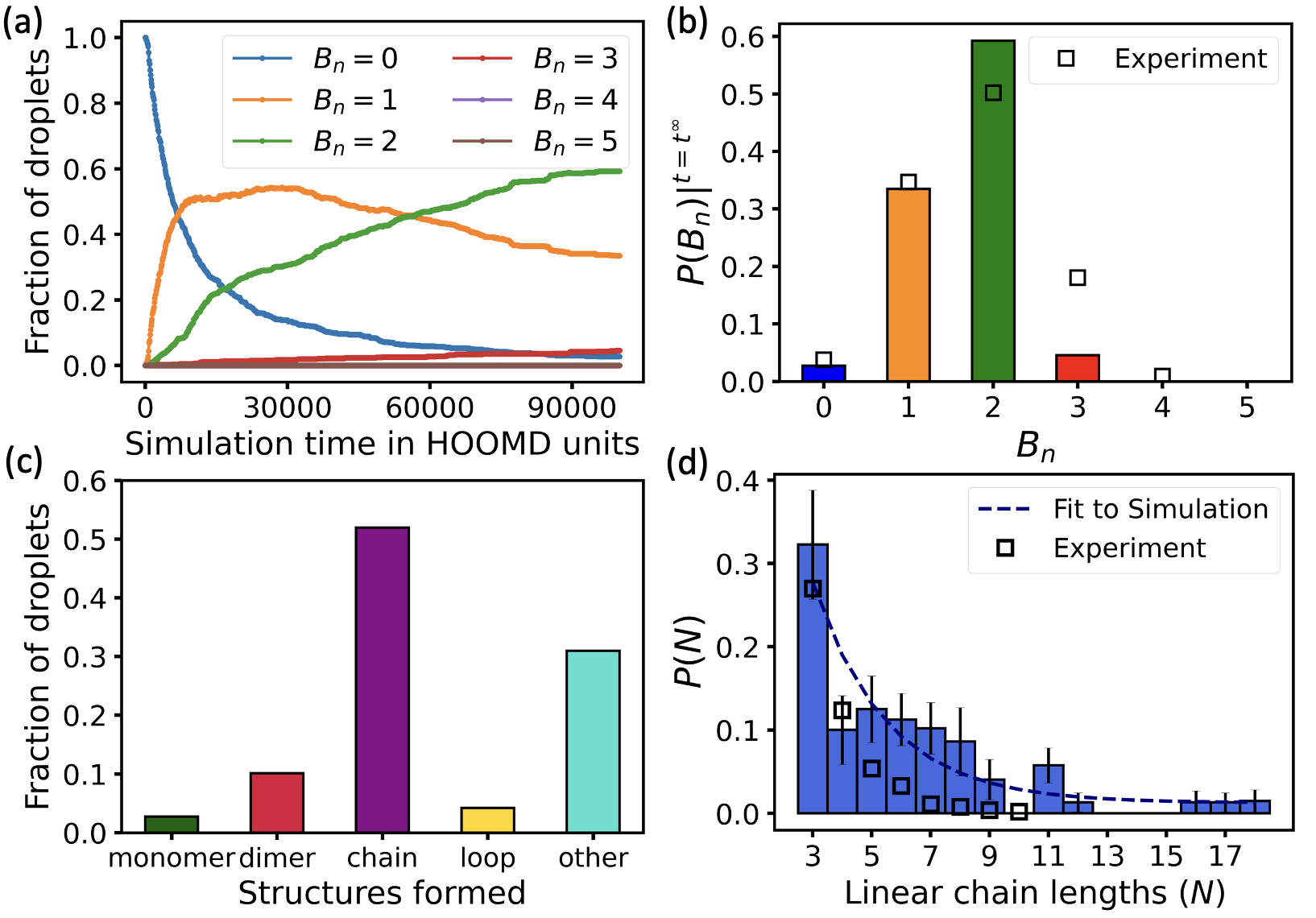}
    \caption{Best choice of parameters for obtaining maximum quality of colloidomer chains: $N_{b} = 100$, $R = 50$, $\phi = 0.3$,  $\varepsilon = 20.7$, $\gamma_A = 1.0$. (a) Fraction of droplets in each possible valence versus time.  (b) Final valence distribution obtained across the 10 simulations. Open squares show optimal valence distribution from Fig. 2g in Ref.~\citenum{mcmullen2018freely}. (c) A  combined histogram showing the fraction of the total number of droplets present as a particular kind of structure from 10 different final configurations for this condition. (d) Distribution of linear chain lengths $P(N)$ obtained for this condition, with the errorbars computed using a bootstrapping procedure described in Sec. \ref{appendix:bootstrapping}. Open squares show experimental length distribution from Fig. 3a in Ref.~\citenum{mcmullen2018freely}. }
    \label{fig:optimize}
\end{figure}

The Einstein relation gives $D=k_\mathrm{B} T/\gamma$, therefore $\tau_\mathrm{collision} \sim \gamma$, the drag on a particle. As described in the simulation methods Sec.~\ref{sec:methods}, $\gamma$ for binder particles is kept small and constant, while $\gamma$ on the droplet particle `A' is varied.
In Fig.~\ref{fig:density_gamma_higheps}, we show that by fixing the previously optimized $R, N_b, \varepsilon$ parameters and varying $\phi$ and $\gamma$, we can maximize the yield of colloidomers. Although we would predict that we would not observe any branching in equilibrium, since droplet collisions are fast in dense suspensions, we do observe many droplets with valence three at the highest initial density. Increasing $\gamma$ slows down the collision rates and therefore improves the yield of linear chains and avoids loop formation (top row); full trajectories for the upper row are also provided in Movie M3.

More quantitatively, we compute the distribution of structures produced at the end of $10^8$ steps, averaging over 10 independent simulations. 
We partition every interconnected assembly using an algorithm described in Sec.~\ref{appendix:clustering}, and then classify these structures as monomers, dimers, linear chains ($N\geq3$), loops, and `other' (at least one droplet has valence $\geq 3$). 
To define the best conditions for chain assembly, we sought to maximize three quantities that are shown in bar charts next to each condition in Fig.~\ref{fig:density_gamma_higheps}: (1) fraction of droplets present in structures that are not monomers or branched---which we call ``success'' (purple bar), (2) fraction of droplets with valence 2 (green bar), and (3) the average maximum chain length (blue bar, which we scale by the highest value $\langle N_{\mathrm{max}} \rangle^{\mathrm{highest}} = 20$ observed for the condition $\phi=0.4$, $\gamma_A=1.0$).

While conditions at $\phi=0.4$ have the longest chains and a high fraction of particles with valence 2, there are also many `errors' due to chain branching, and so we eliminate high density.
Conditions at lower $\phi$ have fewer errors but also shorter chains.
Based on maximizing these three metrics, we choose the condition ($\phi=0.3,\gamma=1.0$) as our best, and we provide more details about the structures observed at this condition in Fig.~\ref{fig:optimize}.   

In Fig.~\ref{fig:optimize}a, we show the evolution of bond valence versus time, which reflects on average, colloidomers are formed by conversion of monomers to dimers, followed by the conversion of dimers to trimers. Fig.~\ref{fig:optimize}b shows the distribution of valence, and for this condition, we observe that only $\approx 5\%$ of droplets had valence higher than 2, which is indicative of a small number of branching points in chains. 
This result actually has a higher yield of linear chains than the best condition found experimentally in Ref.~\citenum{mcmullen2018freely}.
However, even a small fraction of droplets with valence 3 can prevent  extremely high quality assembly of only chains, as shown in Fig.~\ref{fig:optimize}c, where we observe that $\approx 30\%$ of droplets are present in structures  that contain at least one particle of valence 3 (cyan bar), and hence are considered as errors. 
Finally, Fig.~\ref{fig:optimize}d shows that the distribution of chain lengths seems to follow an exponential distribution, with an average length longer than in the optimal conditions in 
Ref.~\citenum{mcmullen2018freely}. An exponential distribution implies that it will be challenging to obtain a large median chain length via a simple self-assembly strategy.
This comports with experimental findings, where for efficiency, a new methodology was developed to engineer longer chains using magnetic fields applied to a dispersion in a ferrofluid \cite{mcmullen2022self}.

\subsection{Application of the model to colloidomer folding}
\label{sec:folding}
Folding of colloidomer chains and the study of their pathways towards stable structures is an area of active research \cite{mcmullen2022self}.
Here, we demonstrate that our CG model can capture experimentally relevant folding behavior of colloidomer chains and use this to highlight additional features of our CG framework.

Our CG framework allows the user to generate an initial linear colloidomer chain of any length and an arbitrary sequence of binder types. 
Moreover, there can be multiple types of binders on the same droplet. 
To study colloidomer folding, we mimic the experimental setup of having two types of binders on the same droplet, C and D, \textit{each of which is self-complementary}. 
Here, C-C bonds are intended to make up the backbone of the chain, while D-D bonds are the ones driving folding. 
As described earlier, our dynamic bonding model can also have temperature dependent binding/unbinding (see Sec.~\ref{appendix:temperature} for full details).
We therefore choose different melting temperatures such that D-D bonds melt at a lower temperature, while C-C bonds stay in place. 

Here, we generate folded structures using a square wave heating and cooling cycle. During the first segment at higher temperature, the backbone adhesion patches form and the chain explores unfolded configurations. After that, repeated cooling and heating are used to generate low energy structures.

In contrast to our work in Sec.~\ref{sec:molecular_optimization} on optimizing colloidomer assembly, here we do want to produce higher valences. This is achieved through the use of smaller $R=20$ and lower $\varepsilon_{DD}=4.6$, where reversible bonding allows for rearrangements such that equilibrated structures form. Low $R$ has the additional benefit of faster folding times, allowing us to generate many structures in relatively little computational time. 
\begin{figure}[ht!]
    \centering
    \includegraphics[width=\columnwidth]{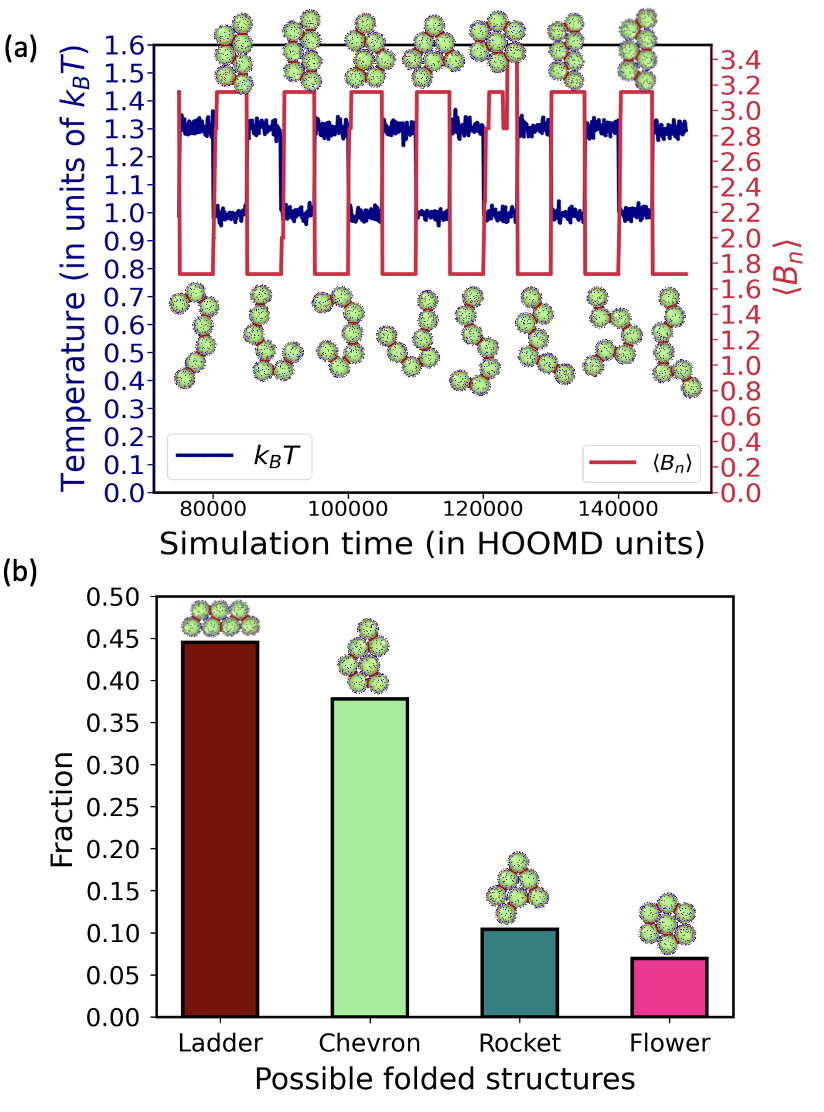}
    \caption{(a) Folding and unfolding cycles shown for a heptamer of droplets with 100 `C' binders and 100 `D' binders on each droplet, $R = 20$, $\varepsilon_{DD} = 4.6$, $\varepsilon_{CC}=\infty$, and $\gamma_{A} = 0.1$. $T_{\mathrm{melt}}=1.2$ for D-D bonds and temperature is cycled between 1.0 and 1.3. 
    The variation of the average bond valence $\langle B_n\rangle$ (red) with simulation time is shown, as the temperature (navy blue) is alternately raised and lowered. An expected $\langle B_n\rangle=1.7$ is obtained when the structures unfold back to chains. (b) Histogram showing the yield of each of the four possible folded rigid structures (ladder, chevron, rocket and flower)\cite{mcmullen2022self} from a total of 1500 folded structures obtained from 300 independent simulations each consisting of 5 folding and unfolding cycles. (only 5/1500 did not reach one of these structures, and are not shown in this histogram).}
    \label{fig:folding_unfolding}
\end{figure}

As a benchmark case to confirm that our model would be applicable in future studies of colloidomer chain folding, we investigate the heptamer case consisting of $N=7$ droplets that has been experimentally realized in Ref.~\citenum{mcmullen2022self}. There are four possible rigid structures which are the low energy states of the heptamer. Scanning a small range of $R$ and $\epsilon_{DD}$ over 15 heating-cooling cycles we uncover the aforementioned condition ($R=20, \varepsilon_{DD}=4.6$) where all four possible structures are observed in a single simulation, as shown in Fig.~\ref{fig:folding_unfolding}a, and Movie M4.

Having obtained all structures in a single long simulation, we wished to quantify the population of each stable folded state. We ran 300 independent simulations each consisting of 5 folding/unfolding cycles and show in  Fig.~\ref{fig:folding_unfolding}b the yields of each structure, which are in good agreement with those reported in  Ref~\citenum{mcmullen2022self}.
The ladder and the chevron structures are
kinetically accessible and have a higher yield 
than the rocket and flower geometries.

From these preliminary studies of folding within our model, we have learned several key principles. 
Firstly, we only observe the flower structure---which has the highest bond count---when $\varepsilon_{DD}$ is very low;this is because folding to this structure is not fully downhill, and requires the breaking of a droplet-droplet bond (dissolving an entire adhesion patch) to reach the final state. 
Second, every droplet contains both C and D binders, and so it could be a concern that the D's become exhausted in forming backbone bonds, which would tend to form faster than bonds farther away in sequence space. In our simulation, this is prevented by choosing $R$ and $N_b$ where the patch is fully saturated by C's (see Fig.~\ref{fig:saturation} and \ref{fig:num_bonds_CC_DD}), meaning that there is no opportunity for D's to enter the backbone.

Moreover, the exclusion of D's from the adhesion patches means there is a smaller area for them to occupy, which results in faster formation of bonds during the folding process.
The factors which contribute to the speed of folding are also important to our results, since both in simulations and in experiments we do not want to wait arbitrarily long times in the cooling phase when generating low energy structures.

\section{Discussion}
In this work, we report a CG model and simulation framework for colloidal liquid droplets with explicit mobile binders. 
The core of this model is a dynamic bonding protocol that satisfies detailed balance, that is very flexible in allowing one to control separately binding and unbinding rate constants, as well as implementing a tunable temperature dependence.
Both the dynamic bonding code and the pyColloidomer framework are easy to use and freely available with examples from \url{https://github.com/hocky-research-group/pyColloidomer_2023}.

Previous modeling works have studied colloidal liquid droplets with implicit mobile linkers, such that bonds are formed or removed based on a statistical mechanical model that predicts the strength of a patch and which can include timescales for bonding.
Our model with explicit binders complements these studies---while having many explicit binders makes the model higher resolution, and hence slower, it also allows us to build insight into the adhesion patch formation process.
For example, the use of explicit binders allowed us to see the effects of excluding particles from patches once they are formed, which had major consequences for ensuring valence=2 structures in optimizing colloidomer assembly, and in preventing binders from being used up in colloidomer backbones in our folding studies.
We also observe that above a certain binding strength ($\varepsilon \approx 10$), the growth of a patch can follow a process characterized by two timescales, where saturation can take much longer than initial formation and recruitment. 
This could have an important effect at higher densities and lower viscosities, where collisions can take place before patches are fully recruited; this effect could be incorporated into simulation models that use a parameterized equation for the recruitment process, such as recently done in Ref.~\citenum{sanchez2022kinetically}.

In ongoing work, we are now using these explicit binders to test the contributions to the free energy of patch formation and patch shape predicted experimentally in Ref.~\citenum{mcmullen2021dna}.
We are also expanding our dynamic bonding model to include the effect of force on unbinding rates \cite{gomez2021molecular} to probe its effect on adhesion patch formation, a dependence which is known to play an important role in the behavior of biomimetic assemblies of cellular adhesion proteins \cite{pontani2012biomimetic}.
Preliminary data shows that our model captures observed behavior for folding of two dimensional colloidomer homopolymer chains; in future work we can use our model to compare the structures and pathways generated through the use of explicit binders with those using very reduced models \cite{trubiano2021thermodynamic,mcmullen2022self}.
We can also trivially expand our folding studies to three dimensions by removing confinement, which will allow us to detail folding pathways in ways that are difficult to quantify in experiment.

Although our model captures what we believe to be the most crucial features, there are simplifications whose effects we would like to investigate in the future. 
For example, the presence of a  spring between the center of the droplet and binders allows the binders' vertical position to vary, and by tuning this parameter we can explore the tendency to form a planar adhesion patch --- however, we are missing the lateral coupling between binders that could be important in the case of deformable droplets. 
Our work also currently employs harmonic springs, and it would be fascinating to investigate the differences where more complex stretching behavior is taken into account. 
Our powerful and flexible framework is freely available and simple to use, and so we hope others will build upon our work and take these studies in new directions.

\section*{Acknowledgements}
We thank Joshua A. Anderson for helping us with technical issues pertaining to our  HOOMD-Blue plugin development. We also thank Stephen Thomas and Eric Jankowski for discussions on the details of their dynamic bonding  plugin.
G.M. and G.M.H. were primarily supported by the National Institutes of Health (NIH) via Award No. R35-GM138312. 
G.M. and D.P. received additional support through the Department of Energy via Award No. DE-SC0019695. A.M. and J.B. were supported by the National Science Foundation under grants, NSF DMR-1710163 and NSF DMR-2105255. J.B.'s work was also supported by the Paris Region (Région Île-de-France) under the Blaise Pascal International Chairs of Excellence.
This work was supported in part through the NYU IT High Performance Computing resources, services, and staff expertise, and simulations were partially executed on resources supported by the Simons Center for Computational Physical Chemistry at NYU (SF Grant No 839534).

\section*{Conflicts of interest}
There are no conflicts to declare.

\clearpage

\setcounter{figure}{0} 
\setcounter{equation}{0}    
\setcounter{section}{0}    

\renewcommand{\thesection}{A\arabic{section}}
\renewcommand{\thefigure}{A\arabic{figure}}
\renewcommand{\thetable}{A\arabic{table}}
\renewcommand{\theequation}{A\arabic{equation}}

\section*{Appendix}

\section{Comparison of our droplet model with experimental geometry}

\begin{figure}[!h]
    \centering
    \includegraphics[width=0.45\textwidth]{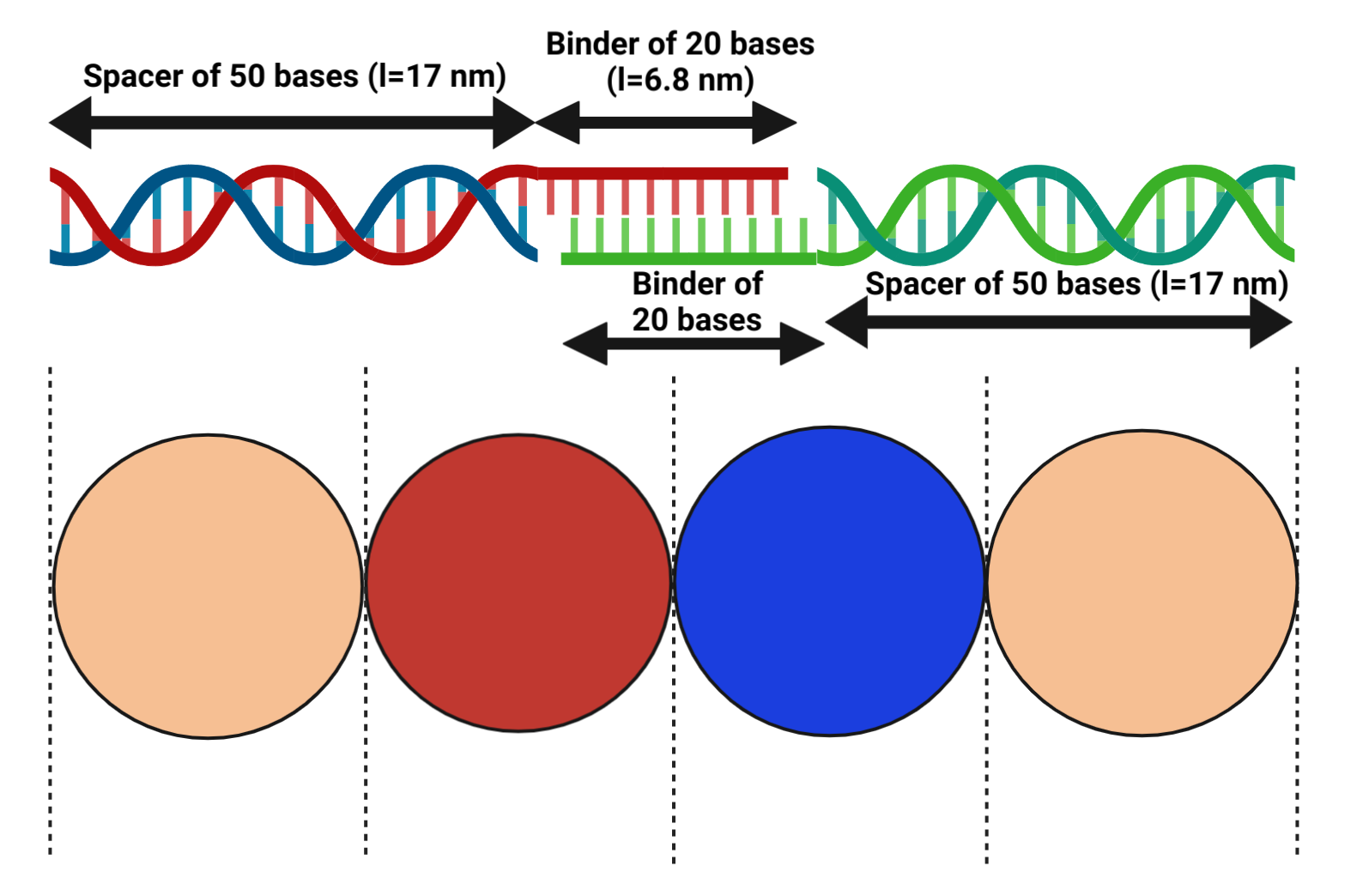}
    \caption{Complementary binders in our CG model mapped onto the dsDNA and ssDNA configuration of Ref.~\citenum{mcmullen2018freely} for scale. Complementary outer binder particles (red and blue) form a dynamic bond representing the interaction between complementary DNA strands (Figure created with \url{https://biorender.com/}).} 
    \label{fig:modeling}
\end{figure}

\section{Non-bonded interaction details}

 \begin{equation}
  U_\mathrm{soft}\big(r\big) = 
  \begin{cases}
                                  \varepsilon_\mathrm{soft} \Bigg[1-\Big(\frac{r}{r_\mathrm{cut}}\Big)^{4}\Bigg] & \text{if $r < r_\mathrm{cut} $} 
                                   \\ 
                                   0 & \text{if $r \geq r_\mathrm{cut}$} \\
  
  \end{cases}
  \label{equation:softV_nosmoothing}
  \end{equation}

\begin{figure}[!ht]
    \centering
    \includegraphics[width=0.4\textwidth]{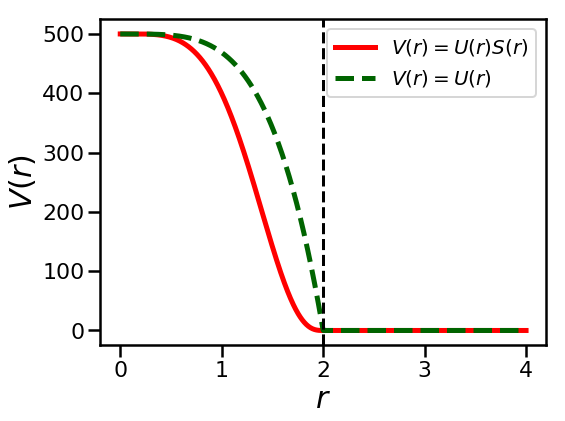}
    \caption{The soft repulsive pair potential $V(r)$ as a function of the distance $r$ between two binder particles of radius $r_{C}=1$. In this figure, $r_\mathrm{cut}=2$ is indicated by the vertical dotted line. The green dotted curve shows the potential $U(r)$ without any smoothing function applied to it and the red solid curve shows the potential $V(r)$ after it is multiplied by a suitable smoothing function $S(r)$.}
    \label{fig:softV}
\end{figure}

 Here, $\varepsilon_\mathrm{soft}$ is the strength of the interaction potential (in units of $k_{\mathrm{B}}T$) and $r_\mathrm{cut}$ is the cut-off distance, as shown by the dotted green curve in Fig.~\ref{fig:softV}. 
 A smoothing function was  applied to this potential $U_\mathrm{soft}(r)$ that results in both the potential energy and the force going smoothly to 0 at $r=r_\mathrm{cut}$, in this case the XLPOR smoothing \cite{anderson2013development} function $S(r)$, given by
\begin{equation}
  S\big(r\big) = 
  \begin{cases}
                                  
                                   1 & \text{if $r< r_\mathrm{on}$} \\
                                   \frac{{(r_\mathrm{cut}^{2}-r^{2}})^2 (r_\mathrm{cut}^{2}+2r^{2}-3r_\mathrm{on}^{2})}{(r_\mathrm{cut}^{2}-r_\mathrm{on}^{2})^{3}} & \text{if $r_\mathrm{on} \leq r \leq r_\mathrm{cut}$} \\
                                   0 & \text{if $r > r_\mathrm{cut}$} \\

  \end{cases}
  \label{equation:smoothingfunc}
  \end{equation}
\par
Here, $r_\mathrm{on}$ is chosen as the point at which the smoothing starts. We set $r_\mathrm{on}=0.1r_\mathrm{cut}$ for our simulations. The modified potential is shown in Fig.~\ref{fig:softV} and is given by 

\begin{equation}
  V_{soft}\big(r\big) = 
  \begin{cases}
                                  S(r) U_\mathrm{soft}(r) & \text{if $r_\mathrm{on} < r_\mathrm{cut} $} 
                                   \\ 
                                   U_\mathrm{soft}(r)-U_\mathrm{soft}(r_\mathrm{cut}) & \text{if $r_\mathrm{on} \geq r_\mathrm{cut}$} \\
  
  \end{cases}
  \label{equation:softV_final}
  \end{equation}
  \par
The soft potential was implemented by using HOOMD-Blue's tabulated potential option (with 1000 interpolation points between $r_\mathrm{min}=0$ and $r_\mathrm{max}=1.05r_\mathrm{cut}$). 

The wall potential is given by a shifted LJ potential with $r_\mathrm{cut}=2^{1/6}(2R)$.
\begin{equation}
 V_\mathrm{wall}\big(r\big)= V_\mathrm{FLJ}\big(r\big) - V_\mathrm{FLJ}\big(r_\mathrm{cut}\big)   
\end{equation}
where, $V_\mathrm{FLJ}(r)$, the force-shifted Lennard-Jones pair potential is given by,
\begin{equation}
V_\mathrm{FLJ}\big(r\big) = 
  \begin{cases}
                                  4\varepsilon_\mathrm{wall} \Big[\big(\frac{\sigma}{r}\big)^{12}-\alpha\big(\frac{\sigma}{r}\big)^{6}\Big] & \text{if $r < r_\mathrm{cut} $} \\ + \Delta V\big(r\big)
                                 \\ \\
                                   0 & \text{if $r \geq r_\mathrm{cut}$} \\
  
  \end{cases}   
\end{equation}
\begin{equation}
\Delta V\big(r\big) = -(r - r_\mathrm{cut}) \frac{\partial V_\mathrm{LJ}}{\partial r}(r_\mathrm{cut})  
\end{equation}
where $\alpha=1$.

\section{Algorithm for binding and unbinding}
\label{appendix:algorithm}
Every $n$ steps of the MD simulation (run using HOOMD-Blue), a Dynamic Bond Updater is called. The Dynamic Bond Updater is an open source C++ plugin (based on previous work on epoxy curing \cite{thomas2018routine}) that stochastically adds or removes dynamic bonds during the course of the MD simulation. If there is more than one dynamic bond type present in the system, the Bond Updater has to be configured for each independently.

Each time the Bond Updater is called, it first iterates over all the existing dynamic bonds to attempt unbinding.
The probability of unbinding is calculated, given by 
\begin{equation}
P_\mathrm{off}^{0} = n\:k_\mathrm{off}\:dt
\end{equation}
where, $dt$ is the timestep of the MD simulation.
For each bond in sequence, a uniform random number $r\in[0,1)$ is generated, and if $r<P_\textrm{off}$ the bond is added to a list of bonds to be removed from the bond table. Once all the possible unbinding events are taken into consideration, we iterate over the list of bonds to be removed and perform unbinding.

After performing unbinding, we create a list of proposed bonds to add. We iterate over particles which were unbound at the start of the binding update (but not those freed by unbinding), and find the closest available complementary particle which is also unbound and whose distance from the particle under consideration is between $l_\textrm{min}$ and $l_\textrm{max}$ (by iterating over HOOMD-Blue's neighbor list).
If an eligible neighbor exists, , and neither particle is already in the proposed bonds list, then this pair is appended.

We then iterate over the proposed bond list, generating new uniform random numbers and creating a bond if $r<P_\textrm{on}$.
By default, 
$P_\mathrm{on}^{0}= n\:k_\textrm{on}\:dt$.
To ensure detailed balance for individual binding reactions, 
the probability of binding is modified such that $P_\mathrm{on}=P_\mathrm{on}^0 e^{-\Delta U(d)/k_\mathrm{B} T}$, where $\Delta U$ is the additional energy added by creating a bond of length $d$, possibly away from its rest length.
As in Ref.~\citenum{maxian2022interplay}, we are putting all of the energetic dependence into the binding step and none in the unbinding step. 
Since we are only using harmonic bonds in this work, 
\begin{equation}
P_\mathrm{on}(d)=P_\mathrm{on}^{0} e^{-k_{\mathrm{dyn}}(d-l_{\mathrm{dyn}})^2/(2k_{\mathrm{B}}T)},
\label{equation:Ponmetropolis}
\end{equation}
where $k_{\mathrm{B}} T$ is the instantaneous temperature of the system.

Detailed balance for the overall binding/unbinding reaction is satisfied to the best of our ability when performing a large set of binding/unbinding reactions at once (as compared to only doing one single binding/unbinding per trial) since every event is independent, and each individually satisfies a Metropolis criterion. 
The only possibly weak breaking of detailed balance comes in the rare situation where upon unbinding, one or both of the particles was assigned a different binding partner since it was not bound to the neighbor from whom its distance is most close to the equilibrium bond length.
In practice, because the configuration evolves $n$ steps between binding/unbinding trials, we do not expect this to cause any substantial non-equilibrium effects. 

\section{Temperature dependence of binding/unbinding}
\label{appendix:temperature}
Our dynamic bonding model allows us to use non-constant values of $k_\textrm{off},k_\textrm{on}$.
As one example, for this work we have incorporated an optional dependence of the rate constants on temperature.
Since our binders here might represent double stranded DNA, which dissociates in a cooperative manner, we implemented an optional tunable sigmoidal dependence on temperature for the rate constants. 
Without trying to match the behavior of any specific module, we implemented a two parameter sigmoidal dependence on temperature to represent cooperative melting,
\begin{equation}
g\big(T\big)=\frac{1}{2}\Bigg[\tanh\Bigg(\alpha\Big(T-T_\mathrm{melt}\Big)\Bigg)+1\Bigg],
\end{equation} 
\begin{equation}
k_\mathrm{on/off}\big(T\big)={k_\mathrm{on/off}}^\mathrm{init} \Big(1-g\big(T\big)\Big) + {k_\mathrm{on/off}}^\mathrm{melt} g\big(T\big)
\label{equation:konkoff}
\end{equation}
where, ${k_\mathrm{on/off}}^\mathrm{melt}$ is the value of the binding (or unbinding) rate constant after the melting of bonds. 

\begin{figure}[!ht]
    \centering
    \includegraphics[width=0.65\columnwidth]{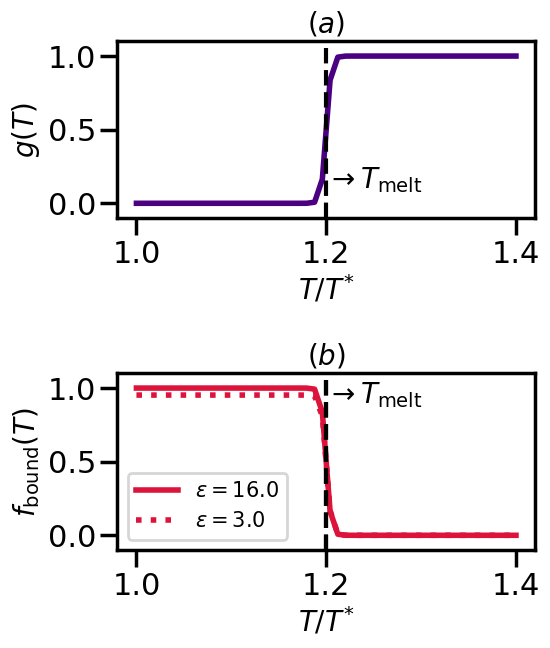}
    \caption{ (a) $g(T)$ vs. $T$ (b) Fraction $f_{\mathrm{bound}}\big(T\big)$ of DNA  that is bound for a single DNA pair vs. $T$ for two different binding strengths $\varepsilon = $ 3.0 and 16.0. (The temperature is in units of $T^{*}$).}
    \label{fig:tempdep}
\end{figure}

Here, we can set the steepness of the transition with the parameter $\alpha$ (which in experiment could be tuned by changing the DNA sequence and sequence length).
$T_\mathrm{melt}$ should be the temperature where the fraction of bonds formed is 0.5.
For a two-state model, the bound fraction is given by 
\begin{equation}
f_{\mathrm{bound}}\big(T\big)=\frac{K_\mathrm{eq}(T)}{1+K_\mathrm{eq}(T)}
\end{equation}
where, $K_\mathrm{eq} (T)  = k_\mathrm{on} (T)/k_\mathrm{off} (T)$.
Ensuring $f_{\mathrm{bound}}(T_\mathrm{melt})=0.5$ requires  $k_\mathrm{on}(T_\mathrm{melt}) = k_\mathrm{off}(T_\mathrm{melt})$. 
Therefore
\begin{equation}
{k_\mathrm{on}}^\mathrm{init} + {k_\mathrm{on}}^\mathrm{melt} = {k_\mathrm{off}}^\mathrm{init} + {k_\mathrm{off}}^\mathrm{melt}
\label{equation:atTmelt}
\end{equation}
since, $g(T_\mathrm{melt})=0.5$.
We choose ${k_\mathrm{on}}^\mathrm{melt} = 0$, so that there is no binding at $T\gg T_\mathrm{melt}$, at which point ${k_\mathrm{off}}^\mathrm{melt}$ can be determined from Eq.~\ref{equation:atTmelt}.
Fig.~\ref{fig:tempdep}(b) shows how $f_{\mathrm{bound}}$ depends on $T_\mathrm{melt}$ using this model.

\bibliography{report} 
\bibliographystyle{rsc} 


\setcounter{figure}{0} 
\setcounter{equation}{0}    
\setcounter{section}{0}    

\renewcommand{\thesection}{S\arabic{section}}
\renewcommand{\thefigure}{S\arabic{figure}}
\renewcommand{\thetable}{S\arabic{table}}
\renewcommand{\theequation}{S\arabic{equation}}

\onecolumn
\begin{center}
    \LARGE Supporting Information
\end{center}
\section{Additional results for Dimers and Trimers}
\subsection{Convergence of the fraction of binders not recruited in an adhesion patch with simulation time}
\label{appendix:convergence}
\begin{figure*}[!ht]
    \centering
    \includegraphics[width=7in]{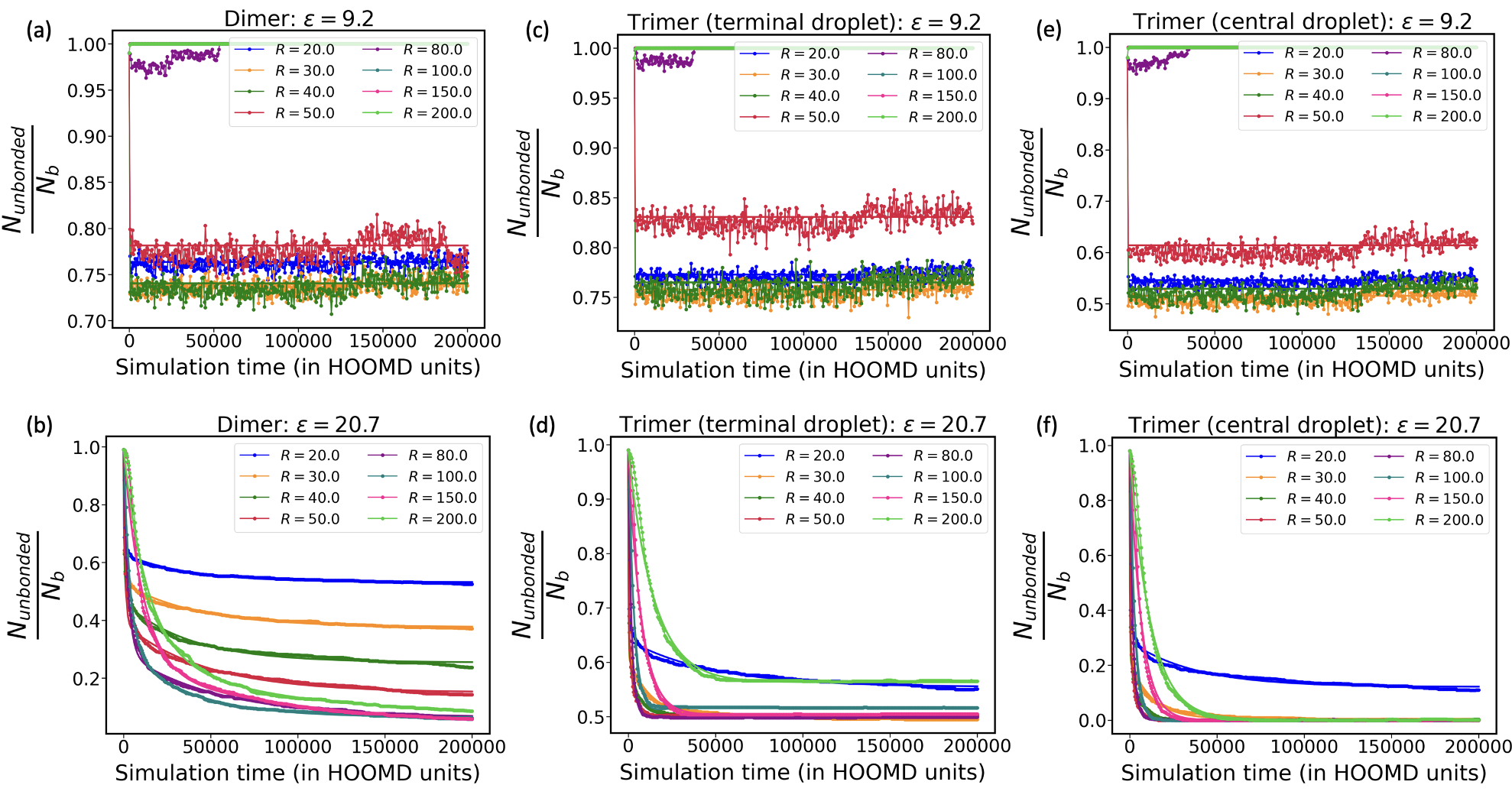}
    \caption{Plots showing the convergence of the fraction of the total binders $N_{b} = 100$ not in any adhesion patch with the simulation time, for two different binding affinities $\varepsilon = 9.2$ (low)---shown in (a),(c),(e) and $\varepsilon = 20.7$ (high)---shown in (b),(d),(f). (a) and (b) show these cases for the dimer, (c) and (d) for the terminal droplet of a trimer, (e) and (f) for the central droplet of the trimer. Each plot shows the convergence for 8 different droplet radii $R$ ranging from 20 to 200. These simulations are all run for $2\times10^{8}$ HOOMD steps which is sufficient for all the individual curves to attain convergence.}
    \label{fig:dimertrimer_convergence}
\end{figure*}
The fraction of the total number of binders not present in the adhesion patch is monitored over time for the case of a dimer, the terminal droplet of a trimer and also its central droplet. For low and high $\varepsilon$, we consider different ways of fitting the curves in order to obtain the converged value of the fraction of free binders.
\par
For binding affinity $\varepsilon \ge 10$, as shown in Fig.~\ref{fig:dimertrimer_convergence}b,d,f for $\varepsilon=20.7$, we fit the fraction of binders not in a patch $f(t)$ to a double exponential function of analytical form $f(t)=(f(0)-a)\exp({-k_{1}t})+(a-b)\exp(-k_{2}t)+b$, where we interpret $k_1$ and $k_2$ to be related to the two different time scales---the recruitment time of the binders and the time taken for the adhesion patch to saturate, explained in Sec.~\ref{sec:dimertrimer}. The fraction of unrecruited binders at saturation of the patch $f(t)|^{t=\infty} = b$ according to this expression. The fitted values for the parameters $a,b,k_1,k_2$ are obtained using the curve fitting feature of SciPy \cite{2020SciPy-NMeth}. From the fitted values of $k_1$ and $k_2$, we can estimate the recruitment time $\tau_1$ and the patch saturation time $\tau_2$, as shown in Tab.~\ref{tab:tabletimescales}.
At larger values of $R>100$, we find that the curves for trimers are well fit by a single exponential, and this emerges naturally in our double exponential fit with $k_1$ and $k_2$ being identical.
\newline

\begin{table}[h!]
\centering
\caption{\textbf{\label{tab:tabletimescales}
A table showing the values of the recruitment time ($\tau_1$) and the adhesion patch saturation time ($\tau_2$) obtained from the double exponential fit for $\varepsilon=20.7$ and $N_{b}=100$}}
\begin{tabular}{l|l|l}
\hline
\textbf{System} & \textbf{$\tau_1 = 1/k_1$} & \textbf{$\tau_2 = 1/k_2$} \\ [1.0 ex]
\hline
Dimer &  &   \\
(i) $R=20.0$  & $1.5\times10^{2}$ & $3.7\times10^{4}$ \\
(ii) $R=30.0$  & $3.2\times10^{2}$ & $4.2\times10^{4}$\\
(iii) $R=40.0$  & $5.3\times10^{2}$ & $3.8\times10^{4}$\\
(iv) $R=50.0$  & $9.3\times10^{2}$ & $4.7\times10^{4}$ \\
(v) $R=80.0$  & $2.5\times10^{3}$  & $5.2\times10^{4}$  \\
(vi) $R=100.0$  & $3.6\times10^{3}$ & $3.6\times10^{4}$ \\
(vii) $R=150.0$  & $1.1\times10^{4}$ & $1.1\times10^{5}$ \\
(viii) $R=200.0$  & $1.3\times10^{4}$ & $7.4\times10^{4}$ \\
\hline
Trimer (terminal droplet) &  &   \\
(i) $R=20.0$  & $1.7\times10^{2}$  & $4.8\times10^{4}$ \\
(ii) $R=30.0$  & $1.7\times10^{2}$ & $1.3\times10^{4}$ \\
(iii) $R=40.0$  & $2.1\times10^{2}$  & $4.7\times10^{3}$ \\
(iv) $R=50.0$  & $2.9\times10^{2}$ & $2.9\times10^{3}$ \\
(v) $R=80.0$  & $1.8\times10^{3}$ & $1.0\times10^{4}$  \\
(vi) $R=100.0$  & $4.1\times10^{3}$ & $5.1\times10^{3}$ \\
(vii) $R=150.0$  & $7.5\times10^{3}$ & - \\
(viii) $R=200.0$  & $1.5\times10^{4}$ & - \\
\hline
Trimer (central droplet) &  &   \\
(i) $R=20.0$  & $1.5\times10^{2}$ & $3.8\times10^{4}$ \\
(ii) $R=30.0$  & $1.7\times10^{2}$ & $1.2\times10^{4}$ \\
(iii) $R=40.0$  & $2.0\times10^{2}$ & $4.6\times10^{3}$ \\
(iv) $R=50.0$  & $3.0\times10^{2}$ & $2.7\times10^{3}$ \\
(v) $R=80.0$  & $1.7\times10^{3}$ & $1.1\times10^{4}$ \\
(vi) $R=100.0$  & $3.0\times10^{3}$ & -  \\
(vii) $R=150.0$  & $6.6\times10^{3}$ & - \\
(viii) $R=200.0$  & $1.2\times10^{4}$  & - \\
\hline
\end{tabular}
\end{table}

\par
For binding affinity $\varepsilon < 10$, as shown in Fig.~\ref{fig:dimertrimer_convergence}a,c,e for $\varepsilon=9.2$, we find that the fraction remains fairly constant over time (for all the droplet radii) with fluctuations characteristic of low $\varepsilon$. For each of these situations, we obtain a mean of this fraction over the last 50\% of the simulation time (in this case, between $10^{8}$ steps and $2\times10^{8}$ steps) and fit the curve to this constant mean value. For all cases the final value of fraction at the final time (at $2\times10^{8}$ steps) has been found to  lie within 5\%  of the fit value. 

\clearpage
\subsection{Saturation of adhesion patch}
\label{appendix:saturation}
\begin{figure}[!ht]
    \centering
    \includegraphics[width=3in]{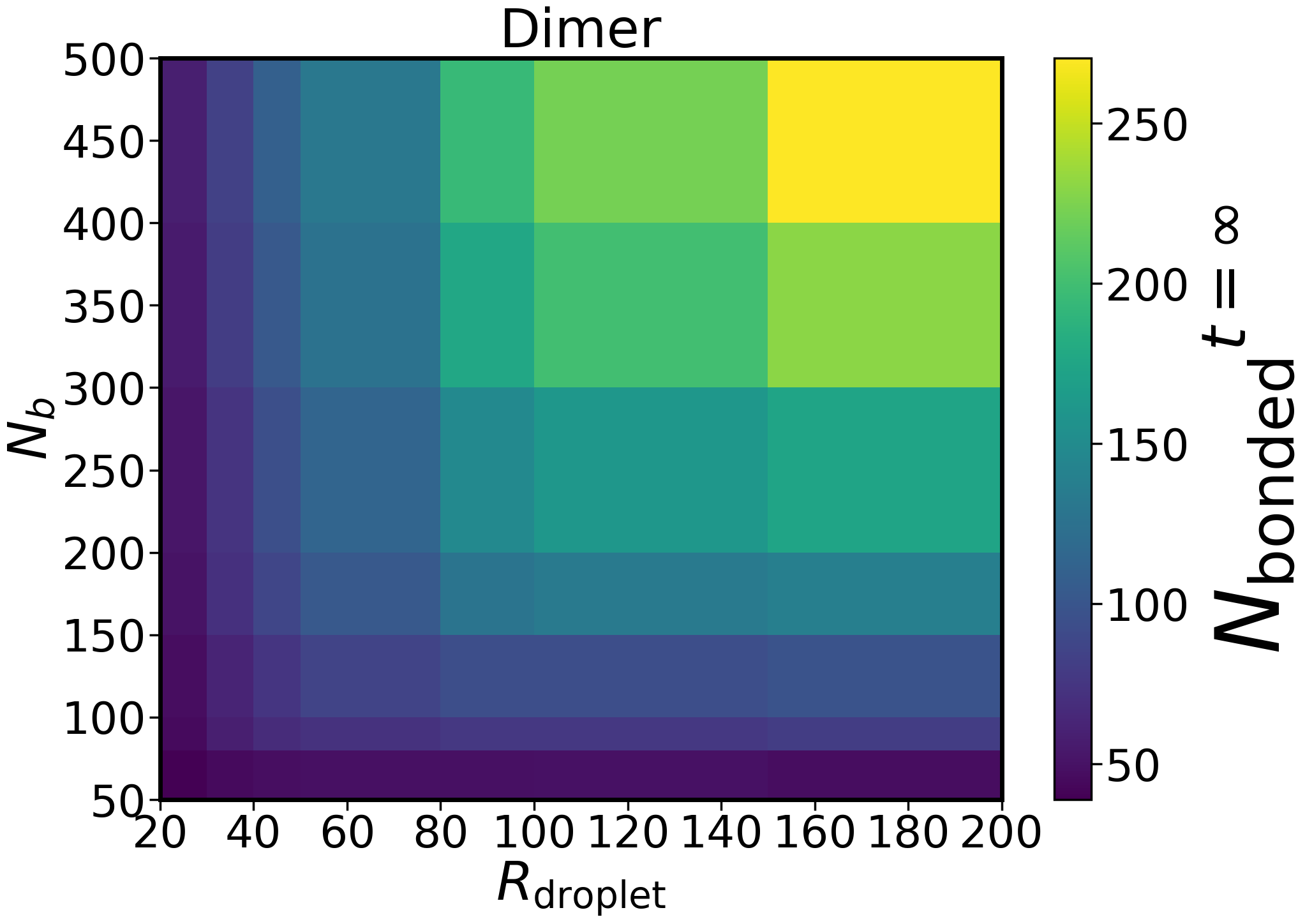}
    \caption{A heat map of the number of binders $N_{b}$ and the droplet radius $R$ showing the number of binders in a saturated adhesion patch (${N_{\textrm{bonded}}}^{t=\infty}$) for high binding affinity ($\varepsilon=20.7$) for a droplet in a dimer.}
    \label{fig:saturation}
\end{figure}

Fig.~\ref{fig:saturation} shows that the number of binders in an adhesion patch saturates due to steric repulsion with $N_\mathrm{bonded}<N_b$.

\section{Additional data and procedures for assembly of 50/50 mixtures of C and D droplets}
\subsection{Clustering via a segmentation algorithm to identify unconnected assemblies and classification of the structures}
\label{appendix:clustering}
For every frame in our simulation, we identify the unconnected assemblies of droplets bonded to each other via a simple segmentation algorithm, using the bond table of droplet pairs. The bond table is a list of unique pairs of droplets that have at least one dynamic bond between them.

From this list of bonded droplet pairs, we want to find the components which have common elements between them. We follow an algorithm to continuously merge sets of pairs that have common elements\footnote{https://stackoverflow.com/questions/4842613/merge-lists-that-share-common-elements}to perform the clustering: (i) We take the first set, say `A' from the list and separate the rest of the list from it. (ii) Next, for every other set B in the list, if B has common element(s) with A then we merge sets A and B and remove B from the list. (iii) This step is repeated until none of the other sets have any overlap with A. (iv) The set A is then added to the output of the clustering. (v) Step (i) is repeated, but now with the rest of the list (excluding `A'). In this way, we ultimately end up with a list of lists where each sub-list consists of droplets which are `bonded' to each other. 

Once the set of unconnected assemblies (or `clusters') is obtained, the next step is to classify them into structures---monomers, dimers, linear chains ($N\geq3$), loops and `other' (which includes any branched colloidomer chains or gels/aggregates with higher droplet valences). 
Monomers are identified as those that do not appear in the bond table.
Next, in order to classify the remaining droplets into the other structures, for every cluster we obtained, we first calculate the valence of each droplet in that cluster from the bond table of droplets, and then count the number of droplets with valence 1 and 2 respectively. If the size of the cluster is 2, then both of the droplets belong to a `dimer'. However, if the size of the cluster list is greater than 2, then the valence information will help us to further identify if it is a linear chain, a loop, or `other'. If the number of droplets in the given cluster with valence 1 is 0 and the number of droplets with valence 2 is equal to the total size of the cluster, then the structure is a `loop'. If the number of droplets with valence 1 is 2 and the number of droplets with valence 2 is equal to the (size of the cluster)-2, then the structure is a `linear colloidomer chain' (for which only the 2 terminal droplets have valence 1 and the rest have valence 2). Else, the cluster is classified as `other'.


\clearpage
\subsection{Distributions of droplet valences, structures and colloidomer chain lengths in the final configurations obtained from 10 independent runs for all choices of $(\phi,\gamma_A)$}
\label{sec:histograms}
   
    \begin{figure*}[!ht]
    \centering
    \includegraphics[width=5.5in]{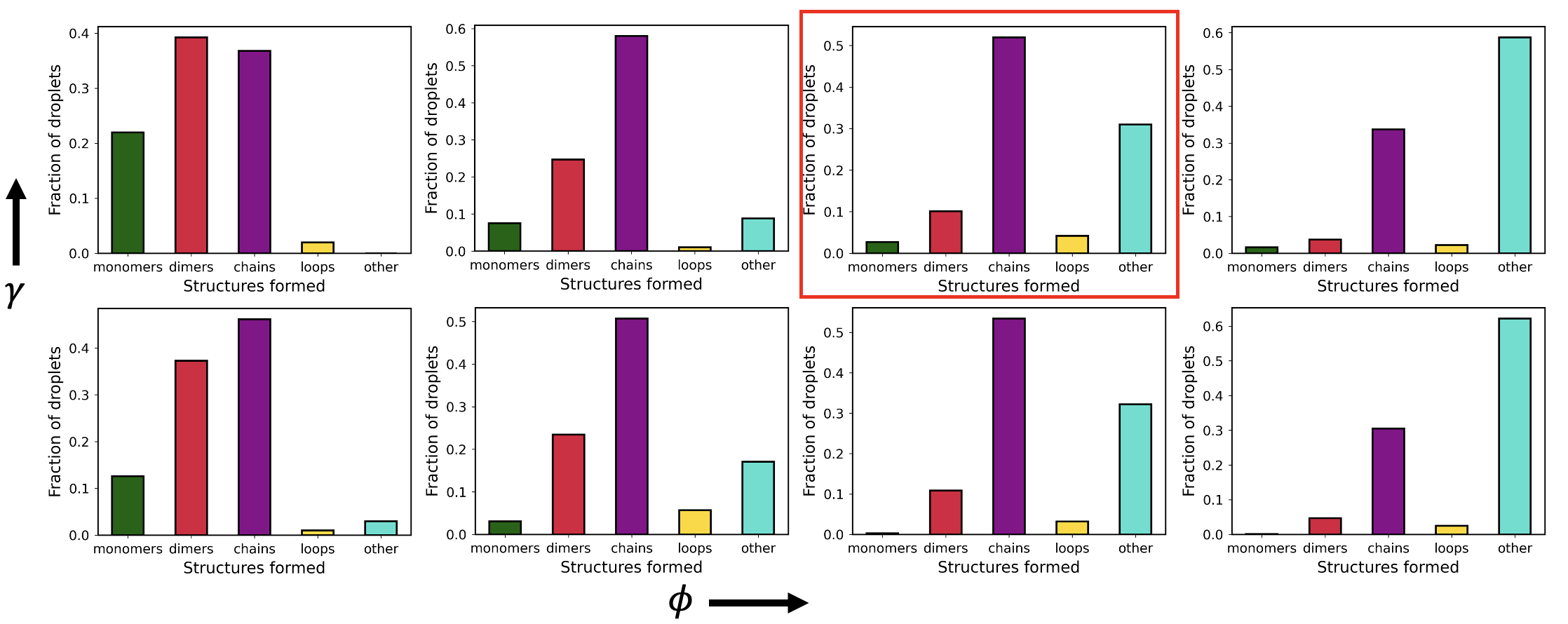}
    \caption{Distributions of structures obtained from all final configurations for each of the $(\phi,\gamma_A)$ pairs, illustrated in Fig.~\ref{fig:density_gamma_higheps}. $\phi = 0.1,0.2,0.3,0.4$ (increasing from left to right) and $\gamma_A = 0.01,1.0$ (bottom and top).}
    \label{fig:histograms_finalstructures}
    \end{figure*}

    \begin{figure*}[!ht]
    \centering
    \includegraphics[width=5.5in]{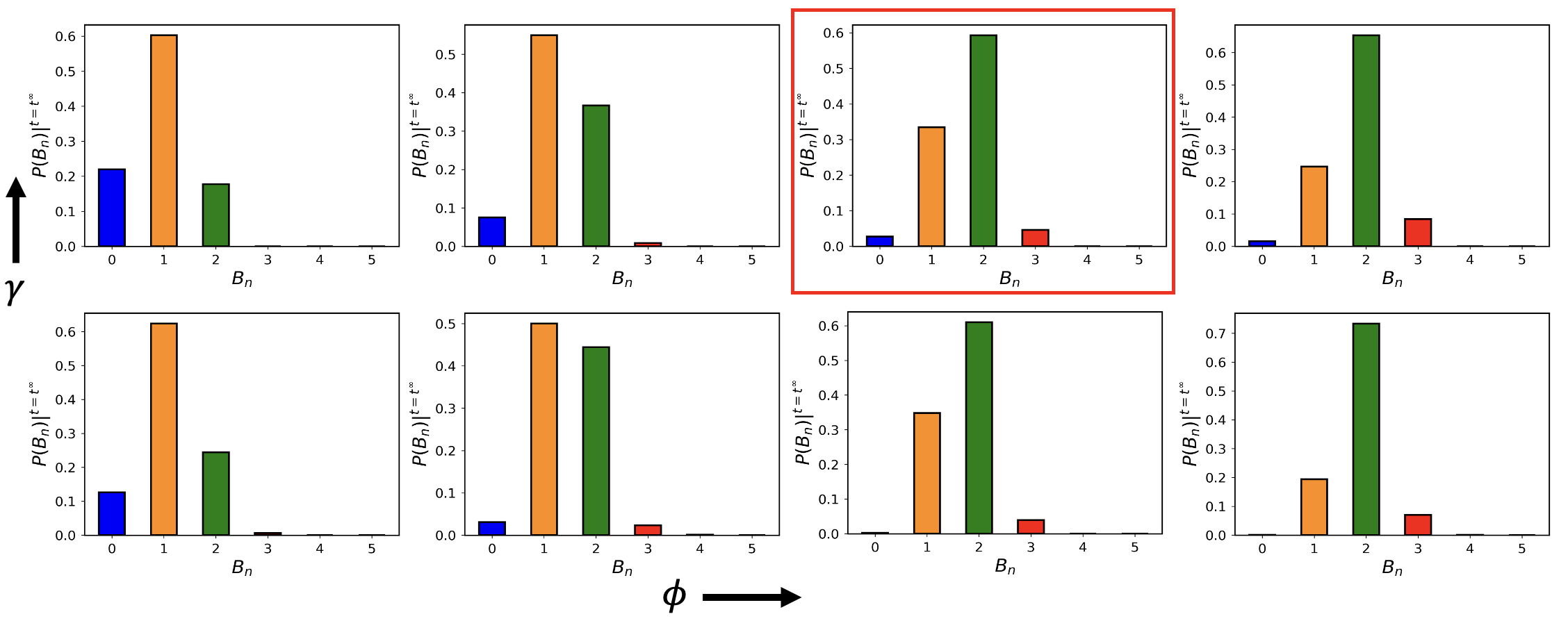}
    \caption{Distributions of droplet valences obtained from all final configurations for each of the $(\phi,\gamma_A)$ pairs, illustrated in Fig.~\ref{fig:density_gamma_higheps}. $\phi = 0.1,0.2,0.3,0.4$ (increasing from left to right) and $\gamma_A = 0.01,1.0$ (bottom and top).}
    \label{fig:histograms_valences}
    \end{figure*}

    \begin{figure*}[!ht]
    \centering
    \includegraphics[width=5.5in]{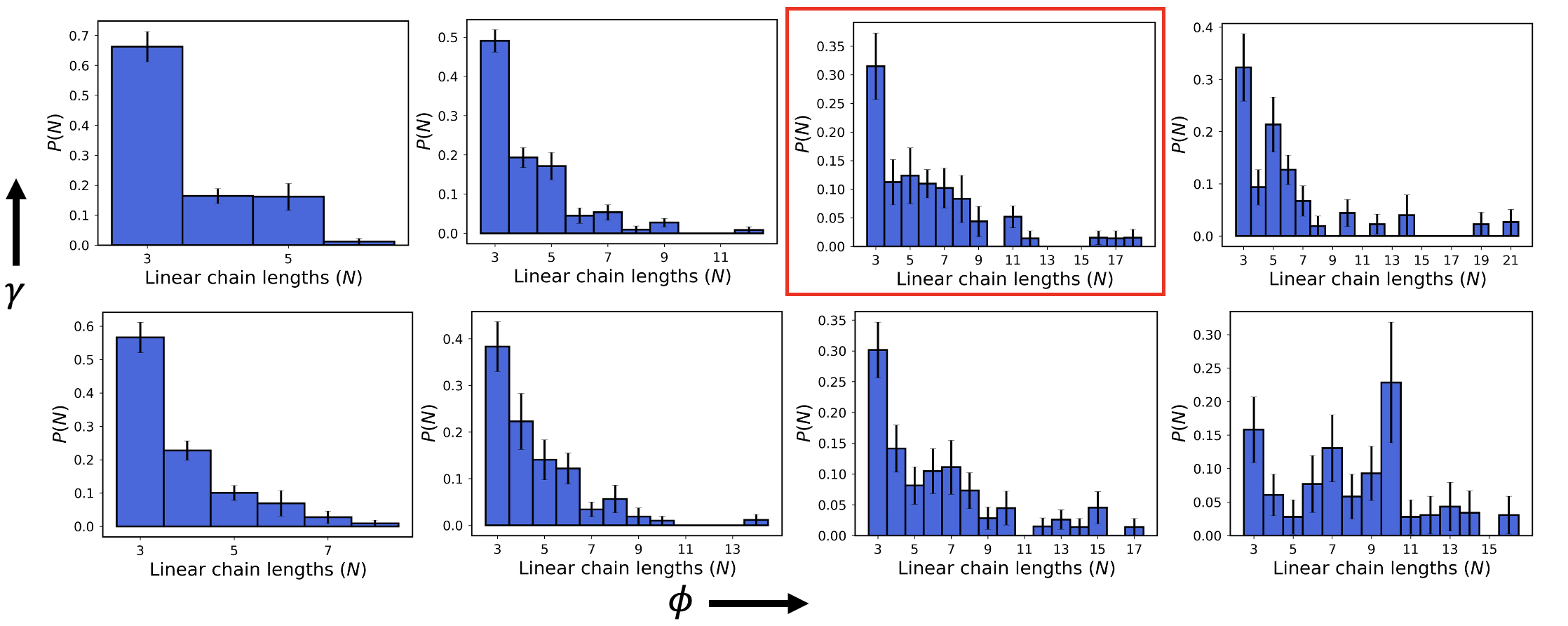}
    \caption{Distributions of colloidomer chain lengths from all final configurations for each of the $(\phi,\gamma_A)$ pairs, illustrated in Fig.~\ref{fig:density_gamma_higheps}. $\phi = 0.1,0.2,0.3,0.4$ (increasing from left to right) and $\gamma_A = 0.01,1.0$ (bottom and top).}
    \label{fig:chainlength_histograms}
    \end{figure*} 
 

\clearpage
\subsection{Effect of kinetic factors on self-assembly in reversible regimes (low and intermediate $\varepsilon$)}
\label{appendix:reversibleregime}
\begin{figure}[!ht]
    \centering
    \includegraphics[width=3.5in]{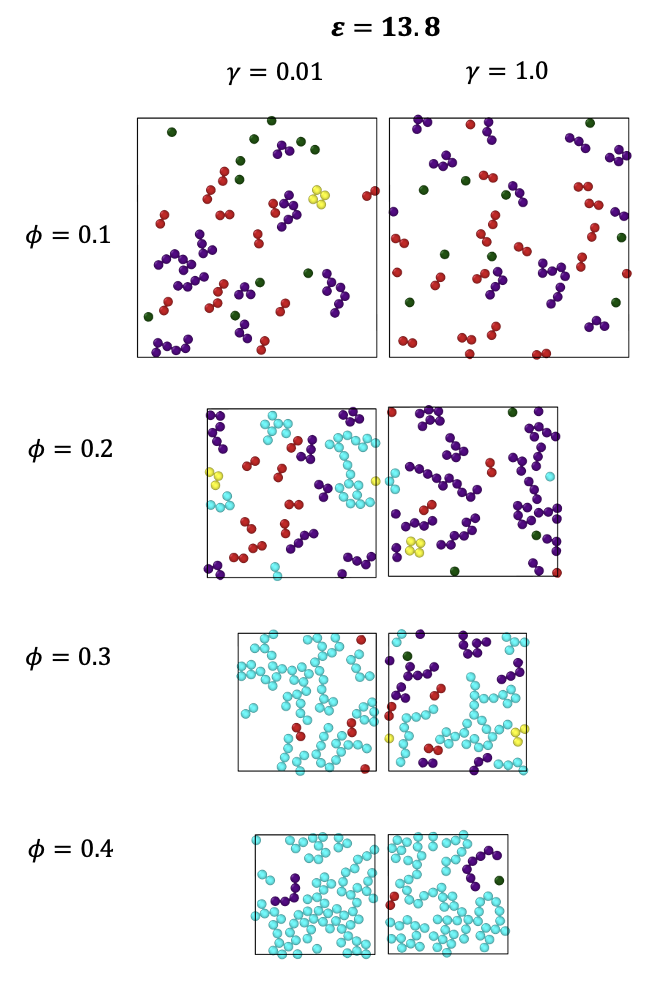}
    \caption{ Effect of $\phi$ and $\gamma_A$ on the kind of self-assembled structures for our system of 81 droplets with $N_{b} = 100$, $R = 50$ for intermediate binding affinity $\varepsilon = 13.8$. The droplets are colored according to the type of structure to which they belong, as explained in Fig.~\ref{fig:density_gamma_higheps}}
    \label{fig:density_gamma_moderateeps}
\end{figure}

\begin{figure}[!ht]
    \centering
    \includegraphics[width=3.5in]{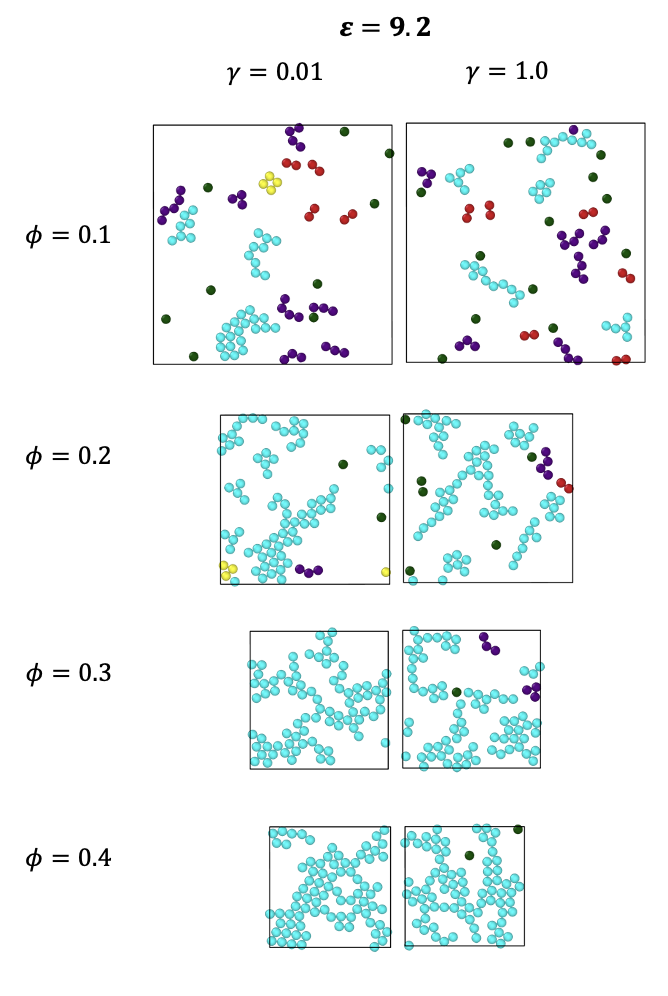}
    \caption{ Effect of $\phi$ and $\gamma_A$ on the kind of self-assembled structures for our system of 81 droplets with $N_{b} = 100$, $R = 50$ for lower binding affinity $\varepsilon = 9.2$. The droplets are colored according to the type of structure to which they belong, as explained in Fig.~\ref{fig:density_gamma_higheps}}
    \label{fig:density_gamma_loweps}
\end{figure}

\begin{figure}[!ht]
    \centering
    \includegraphics[width=3.5in]{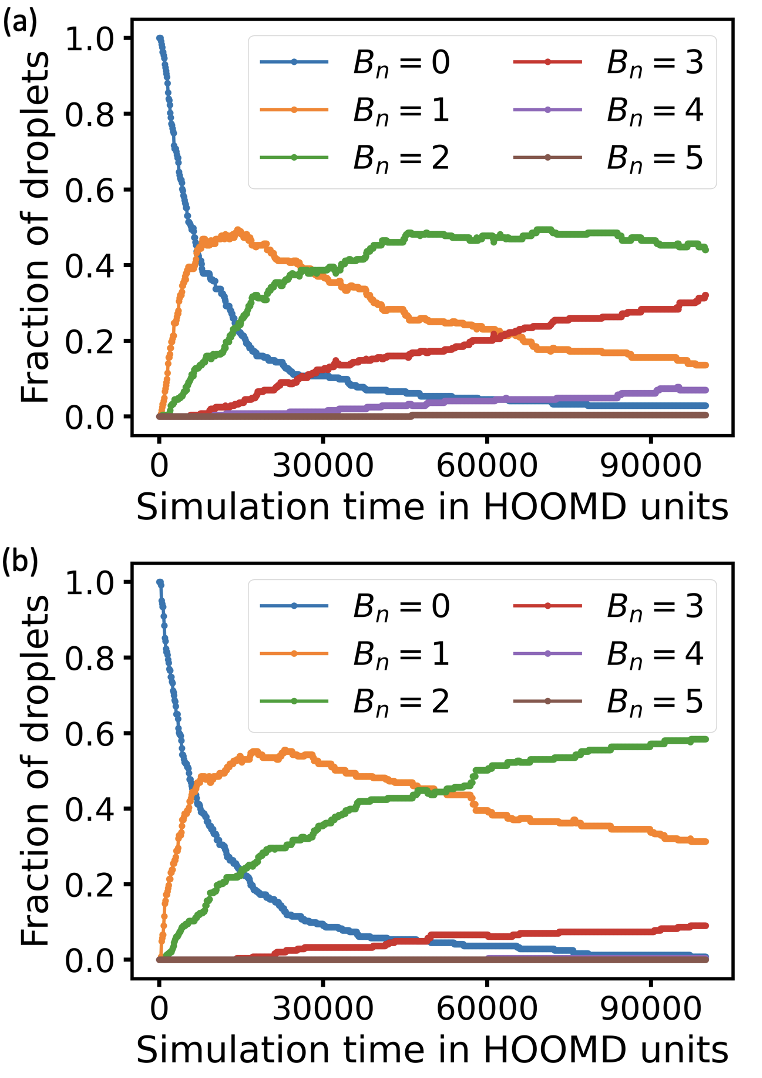}
    \caption{Fraction of droplets with a given bond valence ($B_{n}$) as a function of the simulation time (in HOOMD units) for the optimized condition which gives maximum quality of colloidomer chains at high binding energy: $N_{b} = 100$, $R = 50$, $\phi = 0.3$, $\gamma_{A} = 1.0$ but for (a) lower binding affinity $\varepsilon = 9.2$ and (b) intermediate binding affinity $\varepsilon = 13.8$. The corresponding plot for $\varepsilon=20.7$ is shown in Fig.~\ref{fig:optimize}a.}
    \label{fig:valenceplots_lowerepsilons}
\end{figure}

In Sec.\ref{sec:lattice} we described how kinetic factors such as $\phi$ and $\gamma_{A}$ can dictate the structures formed in self-assembly for a high $\varepsilon$. In this regime, once an adhesion patch forms, binders are very unlikely to redistribute into bonds with other droplets.
However, for reversible binding in case of lower $\epsilon$, we find that this kinetic trapping effect is not observed because redistribution of binders between droplets is allowed here, allowing droplets to potentially achieve their equilibrium valence. 
We show a case of lower ($\varepsilon=9.2$) and intermediate ($\varepsilon=13.8$) binding affinity here to demonstrate the effect of $\phi$ and $\gamma_A$ for reversible binding scenarios (final configurations shown in Figures ~\ref{fig:density_gamma_moderateeps} and \ref{fig:density_gamma_loweps} respectively.). The distribution of bond valences over time for $\phi=0.3$ and $\gamma_{A}=1.0$ is shown in Fig.\ref{fig:valenceplots_lowerepsilons} for lower and intermediate $\varepsilon$. The difference between the two cases is the most profound for lower densities such as $\phi=0.2$, where for $\varepsilon=9.2$, we already obtain aggregates whereas for $\varepsilon=13.8$, we end up mostly in chains, with and without branching. For $\varepsilon=9.2$ and high densities such as $\phi=0.4$, it is almost impossible to differentiate the effect that low and high $\gamma_A$ have, because of system-spanning gels forming in both situations.

\clearpage
\subsection{Bootstrapping procedure}
\label{appendix:bootstrapping}
From the original data set of chain lengths for 10 final configurations, we create 100 new resampled sets of data with replacements (each resampled set is of the same size as the original data set). To create a bootstrap sample, the resampling is done randomly, by choosing a random integer between 0 and (the size of the original data set)-1. The element from the original data set with this random index is then added to the bootstrap sample and this process is repeated for as many times as the size of the original data set. Ultimately from the 100 bootstrap samples obtained, we can calculate the mean fraction of all chain sizes and also the standard deviation. By bootstrapping, we obtain an estimate of the error resulting from having a small number of samples \cite{efron1994introduction}.    

\section{Folding/unfolding simulations}
\subsection{Identification of the folded structures for the heptamer}
\label{appendix:identify_folded}
The folded structures obtained in our 300 independent simulations for the heptamer can be classified as either a ladder, a chevron, a rocket or a flower (see Sec.~\ref{sec:folding}, and Ref.~\citenum{mcmullen2022self}). These structures can be differentiated from one another on the basis of the valence of the droplets present in the structure. A valence list of all the 7 droplets is obtained for every folded structure. The number of droplets with valence 2,3,4,5 and 6 is calculated for each of these structures from the list of valences. We assign the structure based on the counts for each of these valences (see Tab.~\ref{tab:tablevalences7mer}). We found that only 5 of the 1500 folded structures obtained could not be assigned to any of the four folded states mentioned above.

\begin{table}[h!]
\centering
\caption{\textbf{\label{tab:tablevalences7mer}
A table showing the counts of the droplet valences for the different folded geometries for N=7 in two dimensions }}
\begin{tabular}{c|c|c|c|c}
\hline
\textbf{Valence} & \textbf{Ladder} & \textbf{Chevron} & \textbf{Rocket} & \textbf{Flower} \\ [1.0 ex]
\hline
2 & 2 & 2 & 3 & 0 \\
3 & 2 & 3 & 1 & 6 \\
    4 & 3 & 1 & 2 & 0 \\
5 & 0 & 1 & 1 & 0 \\
6 & 0 & 0 & 0 & 1 \\
\hline
\end{tabular}
\end{table}

\subsection{Number of bonds of C-C and D-D types in an adhesion patch \textit{with an adjacent droplet} during folding/unfolding simulations}
\label{appendix:numbonds_foldingsim}
\begin{figure}[!h]
    \centering
    \includegraphics[width=3.5in]{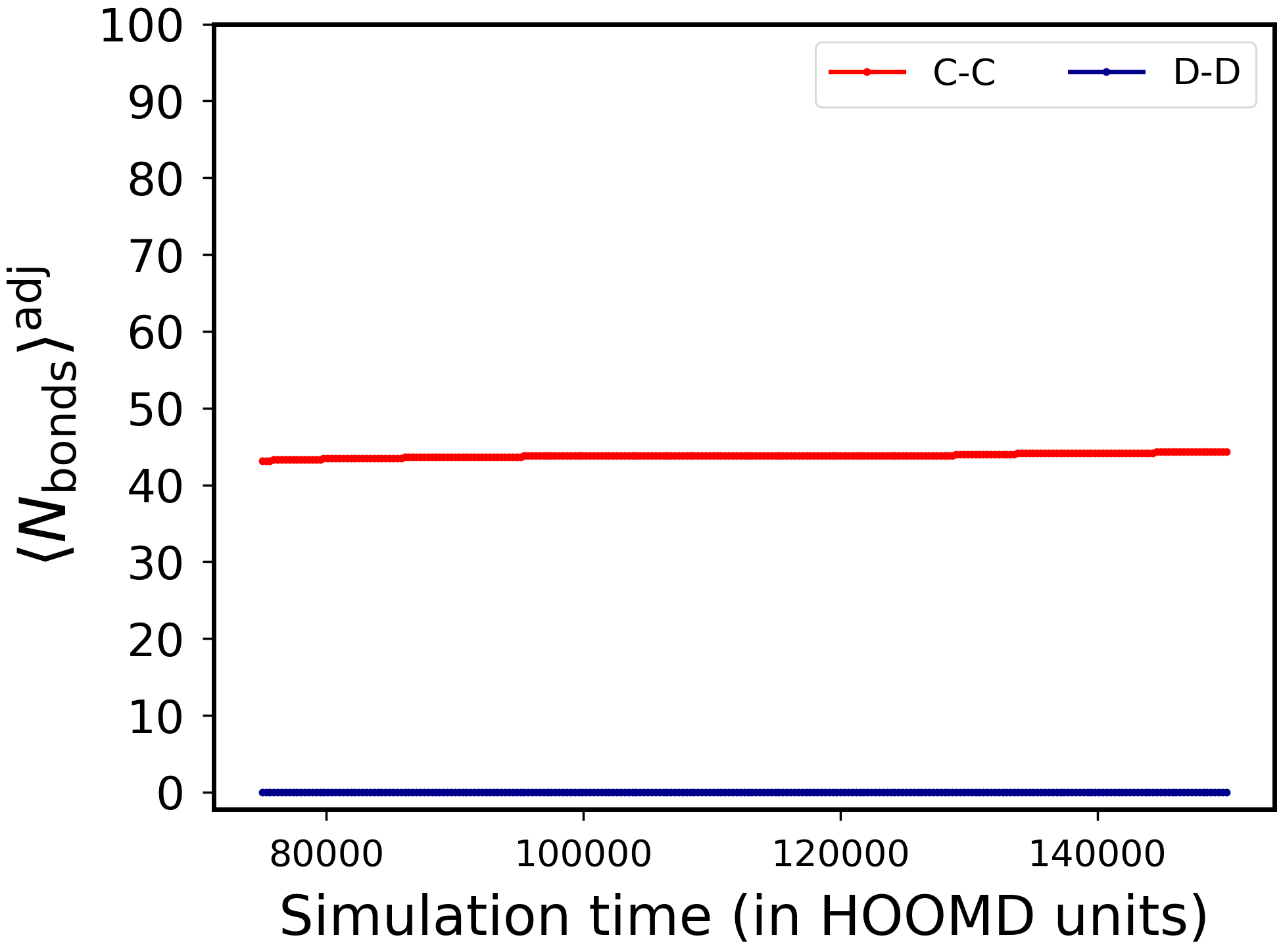}
    \caption{Plot showing the average number of bonds of C-C (backbone) and D-D (secondary) types per adhesion patch between adjacent droplets, as a function of the simulation time for the square wave heating and cooling cycle described in Sec.~\ref{sec:folding}. The conditions for this simulation are: $N_{b} = 200$ ($100$ binders of types C and D respectively on every droplet), $R = 20$, $\gamma_{A} = 0.1$. The choice of $R$ and $N_b$ used here ensures that on an average, $\sim 44\%$ of the C binders get recruited whereas $\sim 0\%$ D binders are able to go into an adhesion patch between any two adjacent droplets. As a result, the adhesion patches between adjacent droplets are always fully saturated by C's (See Sec.\ref{sec:folding})}
    \label{fig:num_bonds_CC_DD}
\end{figure}

\clearpage
\section{Detailed simulation parameters}
\setlength{\tabcolsep}{50pt} 
\renewcommand{\arraystretch}{1.0} 
\begin{table}[h!]
\caption{\textbf{\label{tab:tablegeneral}
A table containing all the general simulation parameters}}
\begin{tabular}{l|l}
\hline
\\
\textbf{Description (Symbol)} & \textbf{Value in HOOMD units} \\ [1.0 ex]
\hline
MD timestep ($dt$)  & 0.0005-0.001   \\
Dimensionality ($d$)  &  2   \\
Temperature ($T$)  & 1.0-1.6   \\
Number of simulation steps run ($n_{\mathrm{steps}}$)  & $10^{8}$ - $2\times 10^{8}$   \\
Radius of droplet ($R$) & 20.0-200.0  \\
Radius of inner binder particle ($r_{B}$) & 1.0  \\
Radius of outer binder particle ($r_{C/D}$)  & 1.0 \\
Number of binders in a droplet ($N_b$) & 50-500 \\
Mass of droplet ($m_{A}$) & 1.0  \\
Mass of inner binder particle ($m_{B}$) & 0.001  \\
Mass of outer binder particle ($m_{C/D}$) & 0.001  \\
Drag coefficient of droplet ($\gamma_{A}$) & 0.01-1.0   \\
Drag coefficient of binder ($\gamma_{\mathrm{binder}}$) & 0.0001   \\
Harmonic bond spring constants:  &    \\
\hspace{2em} (i) $k_{AB}$ & 200.0   \\ 
\hspace{2em} (ii) $k_{BC/BD}$ & 500.0   \\
\hspace{2em} (iii) $k_{ABC/ABD}$ & 10.14  \\
Harmonic bond rest lengths: &    \\
\hspace{2em} (i) $l^{0}_{AB}$ & 50.0 ($R=50.0$)    \\
\hspace{2em} (ii) $l^{0}_{BC/BD}$ & 2.0 ($r_{B}=1.0$)  \\
\hspace{2em} (iii) $\theta^{0}_{ABC/ABD}$ & 3.141593  \\
Epsilon for soft repulsive potential ($\varepsilon_{\mathrm{soft}}$) & 200.0-5000.0  \\
Cut-off distance for soft potential:  &  \\
\hspace{2em} (i) $r_{\mathrm{cut},AA}$ & 110.0 ($R=50.0$)   \\
\hspace{2em} (ii) $r_{\mathrm{cut},AC/AD}$ & 53.0 ($R=50.0$, $r_{C}$=1.0)   \\
\hspace{2em} (iii) $r_{\mathrm{cut},BB}$ & 2.0 ($r_{B}$=1.0)   \\
\hspace{2em} (iv) $r_{\mathrm{cut},CC/DD}$ & 2.0 ($r_{C}$=1.0)   \\
Epsilon for Wall Potential ($\varepsilon_{\mathrm{wall}}$)   &  10.0 ($d=2$)  \\
Cut-off distance for Wall Potential ($r_{\mathrm{cut,wall}}$)  & 112.25 ($R=50.0$)  \\ 
z-coordinate of upper wall plane origin  & 125.0 ($R=50.0$)  \\
z-coordinate of lower wall plane origin  & -125.0 ($R=50.0$)  \\
Initial rate constant for binding ($k_\mathrm{on}^\mathrm{init}$)   & 100.0-200.0   \\
Initial rate constant for unbinding ($k_\mathrm{off}^\mathrm{init}$)   & $10^{-9}$ - 5.0  \\
Rate constant for binding after melting ($k_\mathrm{on}^\mathrm{melt}$)   & 0 \\
Melting Temperature ($T_\mathrm{melt}$)  & 1.2-1.6  \\
Inflexion steepness parameter ($\alpha$) & 200.0 \\
Dynamic bond rest length ($l_{\mathrm{dyn}}$) & 2.0  \\
Dynamic bond spring constant ($k_{\mathrm{dyn}}$) & 10.0  \\
Dynamic bonding minimum distance ($l_{\mathrm{min}}$) & 1.368 ($l_{\mathrm{dyn}}-2\sigma$)\\
Dynamic bonding maximum distance ($l_{\mathrm{max}}$) & 2.632 ($l_{\mathrm{dyn}}+2\sigma$)\\
Dynamic bond checksteps ($n$) & 10   \\
\hline
\end{tabular}
\end{table}

\newpage 
\setlength{\tabcolsep}{38pt} 
\renewcommand{\arraystretch}{1.0} 
\begin{table*}
\caption{\textbf{\label{tab:tabledimertrimer}
A table containing important parameters specific to the simulations for a dimer/trimer of droplets }}
\begin{tabular}{l|l}
\hline
\\
\textbf{Description (Symbol)} & \textbf{Value in HOOMD units} \\ [1.0 ex]
\hline
MD timestep ($dt$)  & 0.001   \\
Temperature ($T$)  & 1.0    \\
Number of simulation steps run ($n_{\mathrm{steps}}$)  &$ 2\times 10^{8} $  \\
Radius of droplet ($R$) & 20.0-200.0  \\
Number of droplets ($N$) & 2,3 \\
Number of binders on each droplet ($N_b$) & 50-500 \\
Drag coefficient of droplet ($\gamma_{A}$) & 0.1   \\
Initial rate constant of binding for CC (${k_\mathrm{on}}^{\mathrm{init},CC}$) & 100.0  \\
Initial rate constant of unbinding for CC (${k_\mathrm{off}}^{\mathrm{init},CC}$) & $10^{-9}$ - 5.0  \\
\hline
\end{tabular}
\end{table*}

\begin{table*}
\caption{\textbf{\label{tab:tablelattice}
A table containing important parameters specific to the self-assembly simulations for a 1:1 mixture of 81 droplets }}
\begin{tabular}{l|l}
\hline
\\
\textbf{Description (Symbol)} & \textbf{Value in HOOMD units} \\ [1.0 ex]
\hline
MD timestep ($dt$)  & 0.001   \\
Temperature ($T$)  & 1.0    \\
Number of simulation steps run ($n_{\mathrm{steps}}$)  & $10^{8}$   \\
Radius of droplet ($R$) & 50.0  \\
Number of droplets ($N$) & 81 ($9\times 9$ lattice) \\
Number of binders on each droplet ($N_b$) & 100 \\
Drag coefficient of droplet ($\gamma_{A}$) & 0.01,1.0   \\
Initial rate constant of binding for CD (${k_\mathrm{on}}^{\mathrm{init},CD}$) & 100.0  \\
Initial rate constant of unbinding for CD (${k_\mathrm{off}}^{\mathrm{init},CD}$) & $10^{-9}$ - 5.0  \\
\hline
\end{tabular}
\end{table*}

\begin{table*}
\caption{\textbf{\label{tab:tablefolding}
A table containing important parameters specific to the folding/unfolding simulations for 300 independent simulations with 5 folding/unfolding cycles in each}}
\begin{tabular}{l|l}
\hline
\\
\textbf{Description (Symbol)} & \textbf{Value in HOOMD units} \\ [1.0 ex]
\hline
MD timestep ($dt$)  & 0.0005   \\
Temperature ($T$)  & 1.0 (quench), 1.3 (heating)   \\
Number of simulation steps run ($n_{\mathrm{steps}}$)  & $10^{8}$ (for 5 folding/unfolding cycles)  \\
Radius of droplet ($R$) & 20.0  \\
Number of droplets ($N$) & 7 \\
Number of `C' type binders on each droplet (${N_{b}}^{C}$) & 100 \\
Number of `D' type binders on each droplet (${N_{b}}^{D}$) & 100 \\
Drag coefficient of droplet ($\gamma_{A}$) & 0.1   \\
Initial rate constant of binding for CC/DD (${k_\mathrm{on}}^{\mathrm{init},CC/DD}$) & 200.0  \\
Initial rate constant of unbinding for CC (${k_\mathrm{off}}^{\mathrm{init},CC}$) & 0  \\
Initial rate constant of unbinding for DD (${k_\mathrm{off}}^{\mathrm{init},DD}$) & 2.0  \\
Melting Temperature for DD ($T_{\textrm{melt},DD}$)  & 1.2  \\
\hline
\end{tabular}
\end{table*}

\clearpage
\section{Additional movies}

M1: Movies showing the growth and saturation of the adhesion patch in a \textit{dimer} for the exact same conditions as illustrated in Fig.~\ref{fig:dimertrimer_pictures}).

\noindent
M2: Movies showing the growth and saturation of the two adhesion patches in a \textit{trimer} for the exact same conditions as illustrated in Fig.~\ref{fig:dimertrimer_pictures}).

\noindent
M3: Movies showing self-assembly of 50/50 mixtures of C and D droplets with $R=50$, $N_b=100$, $\varepsilon=20.7$, $\gamma_{A}=1.0$ for 4 area fractions $\phi=0.1,0.2,0.3,0.4$ (See Fig.~\ref{fig:density_gamma_higheps}).

\noindent
M4: A movie showing 15 heating-cooling cycles for a 7-mer of droplets with $R=20$, $N_b=200$, $\gamma_{A}=0.1$, $\varepsilon_{DD}=4.6$, $\varepsilon_{CC}=\infty$ and $T_{\mathrm{melt},DD}=1.2$ (See Fig.~\ref{fig:folding_unfolding}a).


\end{document}